\documentclass[11pt,twoside,letterpaper]{article} 
\usepackage{times,fancyhdr}
\usepackage[dvips]{graphicx}

\usepackage{epsfig}
\usepackage{amssymb}
\usepackage{amsmath}
\usepackage{amsfonts}
\usepackage{amsthm,amscd}
\usepackage{amsbsy}
\usepackage{latexsym}
\usepackage{bm}
\usepackage{url} 			
\usepackage{layout}
\usepackage{pslatex}
\usepackage{cite}
\usepackage{fleqn} 		
\usepackage{makeidx}
\makeindex 						
\usepackage{layout} 	
\usepackage{epstopdf}	
\usepackage{color}
\usepackage{hyperref}
\usepackage[latin1]{inputenc}
\usepackage[T1]{fontenc}
\usepackage{poligraf} 
\usepackage[letter,cam,center]{crop} 
\usepackage{type1cm} 	
\usepackage{courier}	
\usepackage{lscape} 	


\makeatletter
\def\cleardoublepage{\clearpage\if@twoside \ifodd\c@page\else%
    \hbox{}%
    \thispagestyle{empty}%
    \newpage%
    \if@twocolumn\hbox{}\newpage\fi\fi\fi}
\makeatother

\renewcommand{\thesection}{\arabic{section}.}
\renewcommand{\thesubsection}{\thesection\arabic{subsection}.}

\def\figurename{Figure}
\makeatletter
\renewcommand{\fnum@figure}[1]{\figurename~\thefigure.}
\makeatother

\def\tablename{Table}
\makeatletter
\renewcommand{\fnum@table}[1]{\tablename~\thetable.}
\makeatother

\def\mnras{MNRAS}
\def\apj{ApJ}

\def\apss{Ap\&SS}

\def\pr{Phys. Rev.}

\def\prc{Phys. Rev. C}
\def\prd{Phys. Rev. D}

\def\nat{Nature}

\begin{document}
\title{
{\begin{flushleft}
\vskip 0.45in
{\normalsize\bfseries\textit{Chapter~1}}
\end{flushleft}
\vskip 0.45in
\bfseries\scshape Uniformly rotating neutron stars}}
\author{\bfseries\itshape Kuantay Boshkayev\thanks{E-mail address: kuantay@mail.ru}\\
Institute of Experimental and Theoretical Physics,\\ Faculty of Physics and Technology, Al-Farabi Kazakh National University\\Almaty, Kazakhstan}
\date{}

\maketitle
\thispagestyle{empty}
\setcounter{page}{1}
\thispagestyle{fancy}
\fancyhead{}
\fancyhead[L]{In: Book Title \\
Editor: Boshkayev K., pp. {\thepage-\pageref{lastpage-01}}} 
\fancyhead[R]{ISBN 0000000000  \\
\copyright~2007 Nova Science Publishers, Inc.}
\fancyfoot{}
\renewcommand{\headrulewidth}{0pt}

\begin{abstract}
In this chapter we review the recent results on the equilibrium configurations of static and uniformly rotating neutron stars within the Hartle formalism. We start from the Einstein-Maxwell-Thomas-Fermi equations formulated and extended by Belvedere et al. (2012, 2014). We demonstrate how to conduct numerical integration of these equations for different central densities ${\it \rho}_c$ and  angular velocities $\Omega$ and compute the static $M^{stat}$ and rotating $M^{rot}$ masses, polar $R_p$ and equatorial $R_{\rm eq}$ radii, eccentricity $\epsilon$, moment of inertia $I$, angular momentum $J$, as well as the quadrupole moment $Q$ of the rotating configurations. In order to fulfill the stability criteria of rotating neutron stars we take into considerations the Keplerian mass-shedding limit and the axisymmetric secular instability. Furthermore, we construct the novel mass-radius relations, calculate the maximum mass and minimum rotation periods (maximum frequencies) of neutron stars. Eventually, we compare and contrast our results for the globally and locally neutron star models.
\end{abstract}

\vspace{2in}

\vspace{.08in} \noindent \textbf{Keywords:} Neutron stars, equations of state, mass-radius relation.


\pagestyle{fancy}
\fancyhead{}
\fancyhead[EC]{Kuantay Boshkayev}
\fancyhead[EL,OR]{\thepage}
\fancyhead[OC]{Uniformly rotating neutron stars}
\fancyfoot{}
\renewcommand\headrulewidth{0.5pt}


\section{Introduction}\label{sec:1}

Conventionally, in order to construct the equilibrium configurations of static neutron stars the equations of hydrostatic equilibrium derived by Tolman-Oppenheimer-Volkoff (TOV) \cite{tolman39,oppenheimer39} are widely used. In connection with this, it has been recently revealed in Refs.~\cite{2012NuPhA.883....1B,2011PhLB..701..667R,2011NuPhA.872..286R} that the TOV equations are modified once all fundamental interactions are taken into due account. It has been proposed that the Einstein-Maxwell system of equations coupled with the general relativistic Thomas-Fermi equations of equilibrium have to be used instead. This set of equations is termed as the Einstein-Maxwell-Thomas-Fermi (EMTF) system of equations. Although in the TOV method the condition of local charge neutrality (LCN), $n_e(r)=n_p(r)$ is imposed (see e.g. \cite{haenselbook}), the EMTF method requires the less rigorous condition of global charge neutrality (GCN) as follows
\begin{equation}
\int  \rho_{\rm ch} d^3 r=\int e [n_p(r)- n_e(r)] d^3r = 0,
\end{equation}
where $ \rho_{\rm ch}$ is the electric charge density, $e$ is the fundamental electric charge, $n_p(r)$ and $n_e(r)$ are the proton and electron number densities, respectively. The integration is performed on the entire volume of the system.

The Lagrangian density accounting for the strong, weak, electromagnetic and gravitational interactions consists of the free-fields terms such as the gravitational $\mathcal{L}_g$, the electromagnetic $\mathcal{L}_\gamma$, and the three mesonic fields $\mathcal{L}_\sigma$, $\mathcal{L}_\omega$, $\mathcal{L}_\rho$, the three fermion species (electrons, protons and neutrons) term $\mathcal{L}_f$ and the interacting part in the minimal coupling assumption, $\mathcal{L}_{\rm int}$ given as in Refs.~\cite{2011NuPhA.872..286R,2012NuPhA.883....1B}:
\begin{equation}\label{eq:Lagrangian}
\mathcal{L}=\mathcal{L}_{g}+\mathcal{L}_{f}+\mathcal{L}_{\sigma}+\mathcal{L}_{\omega}+\mathcal{L}_{\rho}+\mathcal{L}_{\gamma}+\mathcal{L}_{\rm int} \;,
\end{equation}
where\footnote{We use the spacetime metric signature (+,-,-,-) and geometric units $G=c=1$ unless otherwise specified.}
\begin{align*}
\mathcal{L}_g &= -\frac{R}{16 \pi},\quad \mathcal{L}_f = \sum_{i=e, N}\bar{\psi}_{i}\left(i \gamma^\mu D_\mu-m_i \right)\psi_i,\\
\mathcal{L}_{\sigma} &= \frac{\nabla_{\mu}\sigma \nabla^{\mu}\sigma}{2}-U(\sigma),\,
\mathcal{L}_{\omega} = -\frac{\Omega_{\mu\nu}\Omega^{\mu\nu}}{4}+\frac{m_{\omega}^{2} \omega_{\mu} \omega^{\mu}}{2},\\
\mathcal{L}_{\rho} &= -\frac{\mathcal{R}_{\mu\nu}\mathcal{R}^{\mu\nu}}{4}+\frac{m_{\rho}^{2} \rho_{\mu} \rho^{\mu}}{2},\quad\mathcal{L}_{\gamma} = -\frac{F_{\mu\nu}F^{\mu\nu}}{16\pi},\\
\mathcal{L}_{\rm int} &= -g_{\sigma} \sigma \bar{\psi}_N \psi_N - g_{\omega} \omega_{\mu} J_{\omega}^{\mu}-g_{\rho}\rho_{\mu}J_{\rho}^{\mu} + e A_{\mu} J_{\gamma,e}^{\mu} -e A_{\mu} J_{\gamma,N}^{\mu}.
\end{align*}
The inclusion of the strong interactions between the nucleons is made through the $\sigma$-$\omega$-$\rho$ nuclear model following Ref.~\cite{boguta77}. Consequently, $\Omega_{\mu\nu}\equiv\partial_{\mu}\omega_{\nu}-\partial_{\nu}\omega_{\mu}$, $\mathcal{R}_{\mu\nu}\equiv\partial_{\mu}\rho_{\nu}-\partial_{\nu}\rho_{\mu}$, $F_{\mu\nu}\equiv\partial_{\mu}A_{\nu}-\partial_{\nu}A_{\mu}$ are the field strength tensors for the $\omega^{\mu}$, $\rho$ and $A^{\mu}$ fields respectively, $\nabla_\mu$ stands for covariant derivative and $R$ is the Ricci scalar. The Lorentz gauge is adopted for the fields $A_\mu$, $\omega_\mu$, and $\rho_\mu$. The self-interaction scalar field potential is $U(\sigma)$, $\psi_N$ is the nucleon isospin doublet, $\psi_e$ is the electronic singlet, $m_i$ stands for the mass of each particle-species and $D_\mu = \partial_\mu + \Gamma_\mu$, where $\Gamma_\mu$ are the Dirac spin connections. The conserved currents are given as $J^{\mu}_{\omega} = \bar{\psi}_N \gamma^{\mu}\psi_N$, $J^{\mu}_{\rho} = \bar{\psi}_N \tau_3\gamma^{\mu}\psi_N$, $J^{\mu}_{\gamma, e} = \bar{\psi}_e \gamma^{\mu}\psi_e$, and $J^{\mu}_{\gamma, N} = \bar{\psi}_N(1/2)(1+\tau_3)\gamma^{\mu}\psi_N$, where $\tau_3$ is the particle isospin. 

In this chapter we adopt the NL3 parameter set \cite{lalazissis97} used in Ref.~\cite{2012NuPhA.883....1B} with $m_\sigma=508.194$ MeV, $m_\omega=782.501$ MeV, $m_\rho=763.000$ MeV, $g_\sigma=10.2170$, $g_\omega=12.8680$, $g_\rho=4.4740$, plus two constants that give the strength of the self-scalar interactions, $g_2=-10.4310$ fm$^{-1}$ and $g_3=-28.8850$.

\begin{figure}[!hbtp]
\centering
\includegraphics[width=0.75\hsize,clip]{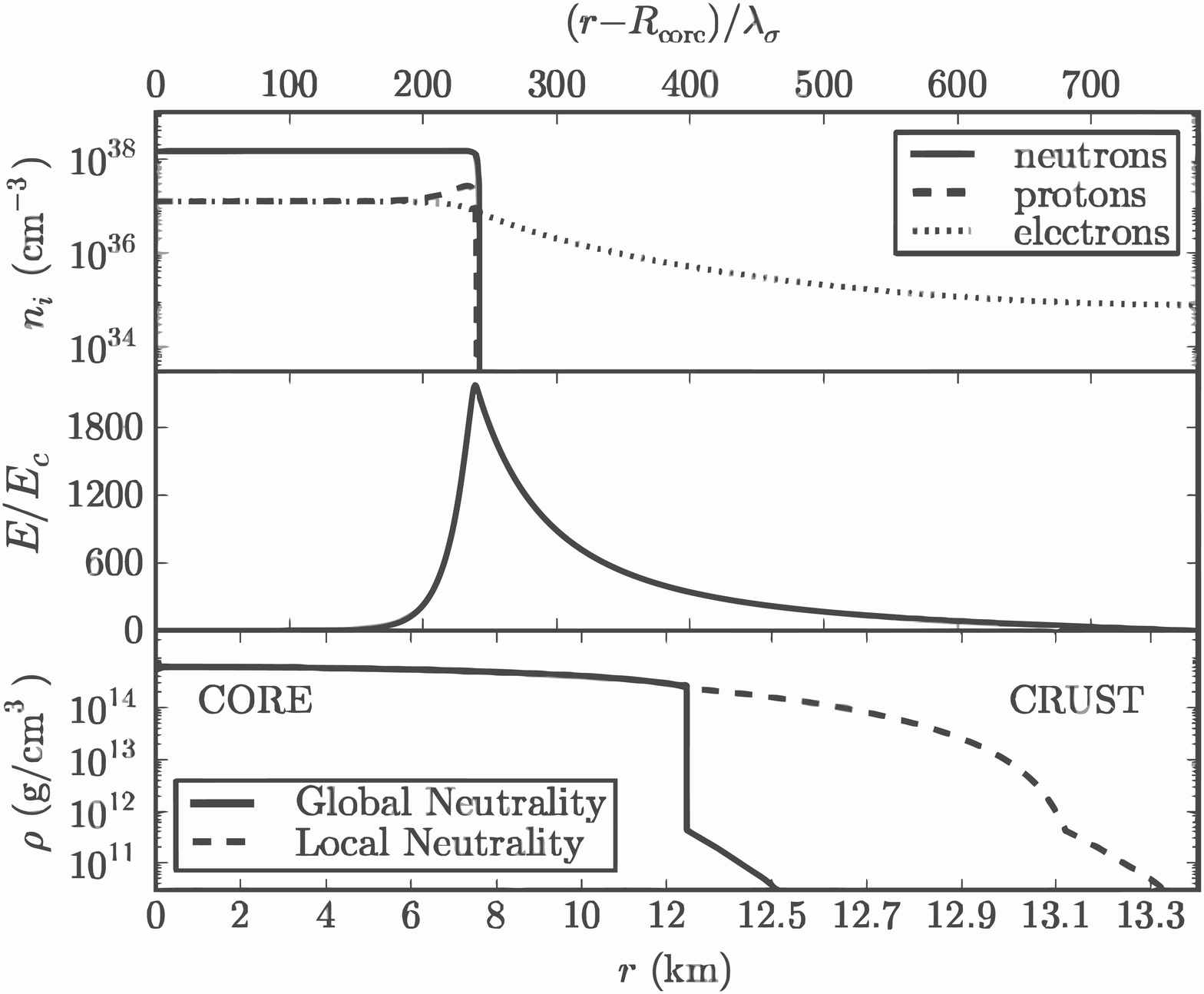} \label{fig:Model}
\caption{The top and middle panels depict the neutron, proton, electron densities and the electric field in units of the critical electric field $E_c$ in the core-crust transition layer, whereas the bottom panel shows a specific example of a density profile inside a neutron star. In this plot we have used for the globally neutral case a density at the edge of the crust equal to the neutron drip density, ${\it\rho}_{\rm drip}\sim 4.3\times 10^{11}$ g cm$^{-3}$.}
\end{figure}

\begin{figure}[!hbtp]
\centering
\includegraphics[width=0.75\hsize,clip]{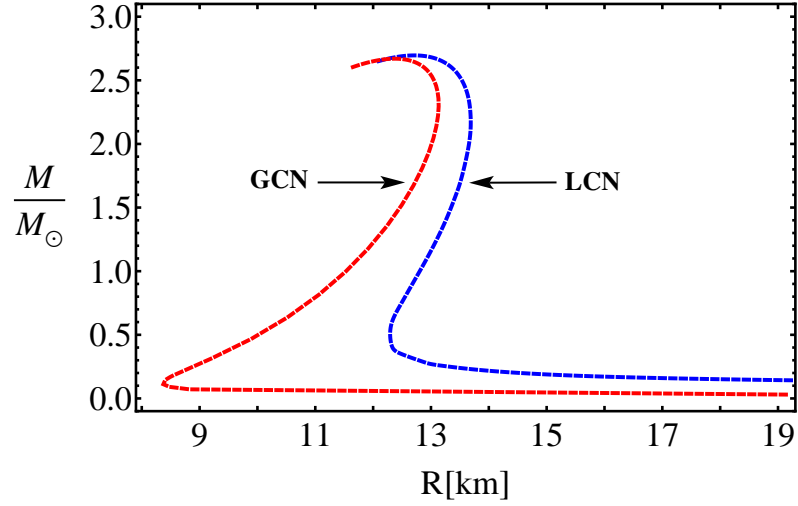}
\caption{Mass-radius relation of the static (non-rotating) neutron stars for both globally and locally neutral configurations. In this plot for the globally neutral case a density at the edge of the crust equal to the neutron drip density, ${\it \rho}_{\rm drip}\sim 4.3\times 10^{11}$ g cm$^{-3}$ has been used. GCN and LCN stand for global and local charge neutrality cases, respectively}\label{fig:MRstat}
\end{figure}
Thus, the system of the EMTF equations \cite{2011NuPhA.872..286R, 2012NuPhA.883....1B, belvedere2014, belvedere2014jkps} is derived from the equations of motion of the above Lagrangian. The solution of the EMTF coupled differential equations gives a novel structure of the neutron star, as shown in Fig~\ref{fig:Model}: a positively charged core at supranuclear densities, ${\it \rho}>{\it \rho}_{\rm nuc}\sim 2.7\times 10^{14}$ g cm$^{-3}$, surrounded by an electron distribution of thickness $\gtrsim \hbar/(m_e c)$, which is negatively charged and a neutral ordinary crust at lower densities ${\it \rho}<{\it \rho}_{\rm nuc}$.

The condition of the thermodynamic equilibrium is given by the constancy of the particle Klein potentials \cite{klein49} extended to account for electrostatic and strong fields \cite{2011PhLB..701..667R,2011NuPhA.872..286R,2012NuPhA.883....1B,belvedere2014jkps}
\begin{equation}\label{eq:klein}
\frac{1}{u^t}\,[\mu_i + (q_i A_\alpha + g_\omega \omega_\alpha  + g_\rho \tau_{3,i} \rho_\alpha) u^\alpha]={\rm constant},
\end{equation}
where the subscript $i$ stands for each kind of particle, $\mu_i$ is the particle chemical potential, $q_i$ is the particle electric charge, $u^t=(g_{tt})^{-1/2}$ is the time component of the fluid four-velocity which satisfies $u_\alpha u^\alpha =1$ and $g_{tt}$ is the t--t component of the spherically symmetric metric. For the static case we have only the time components of the vector fields, $A_0$, $\omega_0$, $\rho_0$.
\begin{equation}\label{eq:metric1}
ds^2=e^{\nu} dt^2-e^{\lambda} dr^2-dr^2-r^2 (d\theta^2+\sin^2\theta d\phi^2)\;.
\end{equation}

The constancy of the Klein potentials (\ref{eq:klein}) leads to a discontinuity in the density at the core-crust transition and, respectively, this generates an overcritical electric field $\sim (m_\pi/m_e)^2 E_c$, where $E_c=m^2_e c^3/(e \hbar)\sim 1.3\times 10^{16}$ Volt cm$^{-1}$, in the core-crust boundary interface. The Klein condition (\ref{eq:klein}) is necessary to satisfy the requirement of thermodynamical equilibrium, together with the Tolman condition (constancy of the gravitationally red-shifted temperature) \cite{1930PhRv...35..904T,klein49}, if finite temperatures are included \cite{2011NuPhA.872..286R}. Particularly, the continuity of the electron Klein potential leads to a decrease of the electron chemical potential $\mu_e$ and density at the core-crust boundary interface. They attain values $\mu^{\rm crust}_e < \mu^{\rm core}_e$ and ${\it \rho}_{\rm crust}<{\it \rho}_{\rm core}$ at the basis of the crust, where global charge neutrality is achieved.

As it has been shown in Refs.~\cite{2012NuPhA.883....1B, belvedere2014}, that the solution of the EMTF equations along with the constancy of the Klein potentials yield a more compact neutron star with a less massive and thiner crust. Correspondingly, this results in a new mass-radius relation which prominently differs from the one given by the solution of the TOV equations with local charge neutrality; see Fig.~\ref{fig:MRstat}.

In this chapter the extension of the previous results obtained in Refs.~\cite{2012NuPhA.883....1B, belvedere2014} are considered. To this end the Hartle formalism \cite{1967ApJ...150.1005H} is utilized to solve the Einstein equations accurately up to second order terms in the angular velocity of the star, $\Omega$ (see section \ref{sec:2}).

For the rotating case, the Klein thermodynamic equilibrium condition has the same form as Eq.~(\ref{eq:klein}), but the fluid inside the star now moves with a four-velocity of a uniformly rotating body, $u^\alpha=(u^t,0,0,u^\phi)$, with (see \cite{HS1967}, for details)
\begin{equation}
u^t=(g_{tt}+2\Omega\,g_{t\phi}+\Omega^2\,g_{\phi \phi})^{-1/2},\qquad u^\phi=\Omega u^t,
\end{equation}
where $\phi$ is the azimuthal angular coordinate and the metric is axially symmetric independent of $\phi$. The components of the metric tensor $g_{\alpha \beta}$ are now given by Eq.~(\ref{eq:rotmetric}) below. It is then evident that in a frame comoving with the rotating star, $u^t=(g_{tt})^{-1/2}$, and the Klein thermodynamic equilibrium condition remains the same as Eq.~(\ref{eq:klein}), as expected.

This chapter is organized as follows: in section~\ref{sec:2} we review the Hartle formalism and consider both interior and exterior solutions. In section~\ref{sec:3} the stability of uniformly rotating neutron stars are explored taking into account the Keplerian mass-shedding limit and the secular axisymmetric instability. In section~\ref{sec:4} the structure of uniformly rotating neutron stars is investigated. We compute there the mass $M$, polar $R_p$ and equatorial $R_{\rm eq}$ radii, and angular momentum $J$, as a function of the central density and the angular velocity $\Omega$ of stable neutron stars both in the globally and locally neutral cases. Based on the criteria of equilibrium we calculate the maximum stable neutron star mass. In section~\ref{sec:5} we construct the new neutron star mass-radius relation. In section~\ref{sec:6} we calculate the moment of inertia as a function of the central density and total mass of the neutron star. The eccentricity $\epsilon$, the rotational to gravitational energy ratio $T/W$, and quadrupole moment $Q$ are shown in section~\ref{sec:7}. The observational constraints on the mass-radius relation are discussed in section \ref{sec:8}. We finally summarize our results in section \ref{sec:9}.

\section{Hartle slow rotation approximation}\label{sec:2}

In his original article, Hartle (1967) \cite{1967ApJ...150.1005H} derived the equilibrium equations of slowly rotating relativistic stars. The solutions of the Einstein equations have been obtained through a perturbation method, expanding the metric functions up to the second order terms in the angular velocity $\Omega$. Under this assumption the structure of compact objects can be approximately described by the total mass $M$, angular momentum $J$ and quadrupole moment $Q$. The slow rotation regime implies that the perturbations owing to the rotation are relatively small with respect to the known non-rotating geometry. The interior solution is derived by solving numerically a system of ordinary differential equations for the perturbation functions. The exterior solution for the vacuum surrounding the star, can be written analytically in terms of $M$, $J$, and $Q$ \cite{1967ApJ...150.1005H,1968ApJ...153..807H}. The numerical values for all the physical quantities are derived by matching the interior and the exterior solution on the surface of the star.

\subsection{The interior Hartle solution}

The spacetime metric for the rotating configuration up to the second order of $\Omega$ is given by \cite{1967ApJ...150.1005H}
\begin{eqnarray}\label{eq:rotmetric}
ds^2 &=& e^{\nu}\left(1+2h\right)dt^2-e^{\lambda}\left[1+\frac{2m}{r-2 M_0}\right]dr^2  \nonumber\\
&&- r^2\left(1+2k\right)\left[d\theta^2+\sin^2\theta\left(d\phi-\omega dt\right)^2\right]+O(\Omega^3) \, ,
\end{eqnarray}
where $\nu=\nu(r)$, $\lambda=\lambda(r)$, and $M_0=M^{J=0}(r)$ are the metric functions and mass profiles of the corresponding seed static star with the same central density as the rotating one; see Eq.~(\ref{eq:metric1}). The functions $h=h(r,\theta)$, $m=m(r,\theta)$, $k=k(r,\theta)$ and the fluid angular velocity in the local inertial frame, $\omega=\omega(r)$, have to be calculated from the Einstein equations. Expanding up to the second order the metric in spherical harmonics we have
\begin{eqnarray}\label{eq:HarmonicExp}
&&h(r,\theta)=h_0(r)+h_2(r)P_2(\cos\theta) \;,\\
&&m(r,\theta)=m_0(r)+m_2(r)P_2(\cos\theta) \;,\\
&&k(r,\theta)=k_0(r)+k_2(r)P_2(\cos\theta) \;,
\end{eqnarray}
where $P_2(cos\theta)$ is the Legendre polynomial of second order. Because the metric does not change under transformations of the type $r\rightarrow f(r)$, we can assume $k_0(r)=0$.

The functions $h=h(r,\theta)$, $m=m(r,\theta)$, $k=k(r,\theta)$ have analytic form in the exterior (vacuum) spacetime and they are shown in the following section. The mass, angular momentum, and quadrupole moment are computed from the matching condition between the interior and exterior metrics.

For rotating configurations the angular momentum is the easiest quantity to compute. To this end we consider only $t,\phi$ component of the Einstein equations. By introducing the angular velocity of the fluid relative to the local inertial frame, $\bar{\omega}(r)=\Omega-\omega(r)$ one can show from the Einstein equations at first order in $\Omega$ that $\bar{\omega}$ satisfies the differential equation
\begin{equation}\label{eq:baromega}
\frac{1}{r^4}\frac{d}{dr}\left( r^4 j \frac{d\bar{\omega}}{dr} \right)+\frac{4}{r}\frac{d j}{dr}\bar{\omega}=0\;,
\end{equation}
where $j(r)=e^{-(\nu+\lambda)/2}$ with $\nu$ and $\lambda$ the metric functions of the seed non-rotating solution (\ref{eq:metric1}).

From the matching conditions, the angular momentum of the star is given by
\begin{equation}\label{eq:J}
J = \frac{1}{6}R^4\left(\frac{d\bar{\omega}}{dr}\right)_{r=R}\;,
\end{equation}
so the angular velocity $\Omega$ is related to the angular momentum as
\begin{equation}\label{eq:Jomega}
\Omega = \bar{\omega}(R)+\frac{2 J}{R^3}\;.
\end{equation}

The total mass of the rotating star, $M$, is given by
\begin{equation}\label{eq:Mrot}
M = M_0+\delta M\;,\qquad \delta M = m_0(R)+J^2/R^3\,,
\end{equation}
where $\delta M$ is the contribution to the mass owing to rotation. The second order functions $m_0$ (the mass perturbation function) and $p_0^*$ (the pressure perturbation function) are computed from the solution of the differential equation
\begin{align}
\frac{d m_0}{dr}&=4\pi r^2 \frac{d{\cal E}}{dP} ({\cal E}+P) p_0^* + \frac{1}{12}j^2 r^4 \left(\frac{d\bar{\omega}}{dr}\right)^2-\frac{1}{3}\frac{dj^2}{dr}r^3 \bar{\omega}^2\;,\\
\frac{d p_0^*}{dr}&=-\frac{m_0 (1+8 \pi r^2 P)}{(r-2 M_0)^2}-\frac{4\pi r^2 ({\cal E}+P)}{(r-2 M_0)}p_0^* 
+ \frac{1}{12}\frac{j^2 r^4}{(r-2 M_0)}\left(\frac{d\bar{\omega}}{dr}\right)^2 \nonumber \\& + \frac{1}{3}\frac{d}{dr} \left(\frac{r^3j^2\bar{\omega}^2}{r-2 M_0}\right)\;,
\end{align}
where ${\cal E}$ and $P$ are the total energy-density and pressure.

Turning to the quadrupole moment of the neutron star, it is given by
\begin{equation}\label{eq:Q}
Q=\frac{J^2}{M_0}+\frac{8}{5}{\cal K} M_0^3\;,
\end{equation}
where ${\cal K}$ is a constant of integration. This constant is fixed from the matching of the second order function $h_2$ obtained in the interior from
\begin{align}
\frac{d k_2}{dr}&=-\frac{d h_2}{dr}-h_2\frac{d\nu}{dr}+\left(\frac{1}{r}+\frac{1}{2}\frac{d\nu}{dr}\right)\bigg[-\frac{1}{3}r^3\bar{\omega}^2\frac{dj^2}{dr} + \frac{1}{6}r^4 j^2 \left(\frac{d\bar{\omega}}{dr}\right)^2\bigg]\;,\\
\frac{d h_2}{dr}&=h_2\bigg\{-\frac{d\nu}{dr}+\frac{r}{r-2 M_0}\left(\frac{d\nu}{dr}\right)^{-1}\bigg[8\pi({\cal E}+P)-\frac{4 M_0}{r^3}\bigg] \bigg\}-\frac{4 (k_2+h_2)}{r (r-2 M_0)}\left(\frac{d\nu}{dr}\right)^{-1}\nonumber\\
&+\frac{1}{6}\bigg[\frac{r}{2}\frac{d\nu}{dr}-\frac{1}{r-2 M_0}\left(\frac{d\nu}{dr}\right)^{-1}\bigg]r^3j^2\left(\frac{d\bar{\omega}}{dr}\right)^2\nonumber\\&-\frac{1}{3}\bigg[\frac{r}{2}\frac{d\nu}{dr}+\frac{1}{r-2 M_0}\left(\frac{d\nu}{dr}\right)^{-1}\bigg]r^2 \bar{\omega}^2\frac{dj^2}{dr}\;,
\end{align}
with its exterior counterpart (see \cite{1967ApJ...150.1005H}).

It is worth emphasizing that the influence of the induced magnetic field owing to the rotation of the charged core of the neutron star in the globally neutral case is negligible \cite{2012IJMPS..12...58B}. In fact, for a rotating neutron star of period $P=10$ ms and radius $R\sim10$ km, the radial component of the magnetic field $B_r$ in the core interior reaches its maximum at the poles with a value $B_r\sim 2.9\times10^{-16}B_c$, where $B_c=m_e^2c^3/(e\hbar)\approx 4.4\times10^{13}$ G is the critical magnetic field for vacuum polarization. The angular component of the magnetic field $B_\theta$, instead, has its maximum value at the equator and, as for the radial component, it is very low in the interior of the neutron star core, i.e. $|B_\theta|\sim 2.9\times 10^{-16}B_c$. In the case of a sharp core-crust transition as the one studied by \cite{2012NuPhA.883....1B} and shown in Fig.~\ref{fig:Model}, this component will grow in the transition layer to values of the order of $|B_\theta|\sim 10^2 B_c$ \cite{2012IJMPS..12...58B}. However, since we are here interested in the macroscopic properties of the neutron star, we can ignore at first approximation the presence of electromagnetic fields in the macroscopic regions where they are indeed very small, and safely apply the original Hartle formulation without any generalization.


\subsection{The exterior Hartle solution}\label{app:1a}

In this subsection we consider the exterior Hartle solution though in the literature it is widely known as the Hartle-Thorne solution. One can write the line element given by eq.~(\ref{eq:rotmetric}) in an analytic closed-form outside the source as function of the total mass $M$, angular momentum $J$, and quadrupole moment $Q$ of the rotating star. The angular momentum $J$ along with the angular velocity of local inertial frames $\omega(r)$, proportional to $\Omega$, and the functions $h_0$, $h_2$, $m_0$, $m_2$, $k_2$, proportional to $\Omega^2$, are derived from the Einstein equations (for more details see \cite{1967ApJ...150.1005H,1968ApJ...153..807H}).  Following this prescriptions the Eq.~\ref{eq:rotmetric} becomes:
\begin{align}\label{ht1}
ds^2&=\left(1-\frac{2{ M }}{r}\right)\bigg[1+2k_1P_2(\cos\theta)
+2\left(1-\frac{2{ M}}{r}\right)^{-1}\frac{J^{2}}{r^{4}}(2\cos^2\theta-1)\bigg]dt^2\nonumber\\
&+\frac{4J}{r}\sin^2\theta dt d\phi-\left(1-\frac{2{ M}}{r}\right)^{-1}\times\bigg[1-2\left(k_1-\frac{6 J^{2}}{r^4}\right)P_2(\cos\theta) \nonumber \\
&-2\left(1-\frac{2{ M}}{r}\right)^{-1}\frac{J^{2}}{r^4}\bigg]dr^2-r^2[1-2k_2P_2(\cos\theta)](d\theta^2+\sin^2\theta d\phi^2),
\end{align}
where
\begin{align*}
k_1&=\frac{J^2}{M r^3}\left(1+\frac{M}{r}\right)+\frac{5}{8}\frac{Q-J^{2}/{M}}{M^3}Q_2^2(x) \;,\\
k_2&=k_1+\frac{J^{2}}{r^4}+\frac{5}{4}\frac{Q-J^{2}/{ M}}{{ M}^2r \sqrt{1-2M/r}}Q_2^1(x) \;,
\end{align*}
and
\begin{align*}
\label{legfunc}
Q_2^1(x)&=(x^2-1)^{1/2}\left[\frac{3x}{2}\ln\left(\frac{x+1}{x-1}\right)-\frac{3x^2-2}{x^2-1}\right] \;, \\
Q_2^2(x)&=(x^2-1)\left[\frac{3}{2}\ln\left(\frac{x+1}{x-1}\right)-\frac{3x^3-5x}{(x^2-1)^2}\right] \;,
\end{align*}
are the associated Legendre functions of the second kind, being $P_2(\cos\theta)=(1/2)(3\cos^2\theta-1)$ the Legendre polynomial, and $x=r/M -1$. This form of the metric is known in the literature as the Hartle-Thorne metric. To obtain the exact numerical values of $M$, $J$ and $Q$, the exterior and interior line elements have to be matched at the surface of the star. It is worth noticing that in the terms involving $J^2$ and $Q$, the total mass $M$ can be directly substituted by $M_0=M^{J=0}$ since $\delta M$ is already a second order term in the angular velocity.

\section{Stability of uniformly rotating neutron stars}\label{sec:3}

\subsection{Secular axisymmetric instability}\label{subsec:3.1}

In a sequence of increasing central density in the $M$-${\it \rho}_c$ curve, ${\it \rho}_c\equiv {\it \rho}(0)$, the maximum mass of a static neutron star is determined as the first maximum of such a curve, namely the point where $\partial M$/$\partial {\it \rho}_c=0$. This derivative establishes the axisymmetric secular instability point, and if the perturbation obeys the same equation of state (EOS) as the equilibrium configuration, it coincides also with the dynamical instability point (see e.g. Ref.~\cite{shapirobook}). In the rotating case, the situation becomes more complicated and in order to find the axisymmetric dynamical instability points, the perturbed solutions with zero frequency modes (the so-called neutral frequency line) have to be calculated. Friedman et al. (1988) \cite{1988ApJ...325..722F} however, following the works of Sorkin (1981, 1982) \cite{1981ApJ...249..254S,1982ApJ...257..847S}, described a turning-point method to obtain the points at which secular instability is reached by uniformly rotating stars. In a constant angular momentum sequence, the turning point is located in the maximum of the mass-central density relation, namely the onset of secular axisymmetric instability is given by
\begin{equation}\label{eq:TurningPoint}
\left[\frac{\partial M\left({\it \rho}_c,J\right)}{\partial{\it \rho}_c}\right]_{J=\rm constant}=0 \;,
\end{equation}
and once the secular instability sets in, the star evolves quasi-stationarily until it reaches a point of dynamical instability where gravitational collapse sets in (see e.g. \cite{2003LRR.....6....3S}).

The above equation determines an upper limit for the mass at a given angular momentum $J$ for a uniformly rotating star, however this criterion is a sufficient but not necessary condition for the instability. This means that all the configurations with the given angular momentum $J$ on the right side of the turning point defined by Eq.~(\ref{eq:TurningPoint}) are secularly unstable, but it does not imply that the configurations on the left side of it are stable. An example of dynamically unstable configurations on the left side of the turning-point limiting boundary in neutron stars was recently shown in Ref.~\cite{2011MNRAS.416L...1T}, for a specific EOS.

In order to investigate the secular instability of uniformly rotating stars one should select fixed values for the angular momentum. Then construct mass-central density or mass-radius relations. From here one has to calculate the maximum mass and that will be the turning point for a given angular momentum. For different angular momentum their will be different maximum masses. By joining all the turning points together one obtains axisymmetric secular instability line (boundary). This boundary is essential for the construction of the stability region for uniformly rotating neutron stars (see next sections and figures).
\subsection{Keplerian mass-shedding instability and orbital angular velocity of test particles}\label{subsec:3.2}

The maximum velocity for a test particle to remain in equilibrium on the equator of a star, kept bound by the balance between gravitational and centrifugal force, is the Keplerian velocity of a free particle computed at the same location. As shown, for instance in \cite{2003LRR.....6....3S}, a star rotating at Keplerian rate becomes unstable due to the loss of mass from its surface. The mass shedding limiting angular velocity of a rotating star is the Keplerian angular velocity evaluated at the equator, $r=R_{\rm eq}$, i.e. $\Omega_K^{J\neq0}=\Omega_K(r=R_{\rm eq})$. Friedman (1986) \cite{Friedman1986} introduced a method to obtain the maximum possible angular velocity of the star before reaching the mass-shedding limit; however \cite{2008AcA....58....1T} and \cite{BBRS2013}, showed a simpler way to compute the Keplerian angular velocity of a rotating star. They showed that the mass-shedding angular velocity, $\Omega_K^{J\neq0}$, can be computed as the orbital angular velocity of a test particle in the external field of the star and corotating with it on its equatorial plane at the distance $r=R_{\rm eq}$.

It is possible to obtain the analytical expression for the angular velocity $\Omega$ given by Eq.~(\ref{eq:omegaKep}) with respect to an observer at infinity, taking into account the parameterization of the four-velocity $u$ of a test particle on a circular orbit  in equatorial plane of axisymmetric stationary spacetime, regarding as parameter the angular velocity $\Omega$ itself:
\begin{equation}
u=\Gamma[\partial_t+\Omega\partial_{\phi}] \;,
\end{equation}
where $\Gamma$ is a normalization factor such that $u^{\alpha}u_{\alpha}=1$. Normalizing and applying the geodesics conditions we get the following expressions for $\Gamma$ and $\Omega=u^{\phi}/u^{t}$
\begin{equation}
\label{eight}
\Gamma=\pm(g_{tt}+2\Omega g_{t\phi}+\Omega^2 g_{\phi\phi})^{-1/2}\;, \quad g_{tt,r}+2\Omega g_{t\phi,r}+\Omega^2 g_{\phi\phi,r}=0 \;.
\end{equation}
Thus, the solution of the system of Eq.~(\ref{eight}) can be written as
\begin{equation}
\Omega^\pm_{\rm orb}(r)=\frac{u^{\phi}}{u^{t}}=\frac{-g_{t\phi,r}\pm\sqrt{(g_{t\phi,r})^2-g_{tt,r}g_{\phi\phi,r}}}{g_{\phi\phi,r}} \;,
\end{equation}
where $+/-$ stands for co-rotating/counter-rotating orbits, $u^{\phi}$ and $u^{t}$ are the angular and time components of the four-velocity respectively, and a colon stands for partial derivative with respect to the corresponding coordinate. To determine the mass shedding angular velocity (the Keplerian angular velocity) of the neutron stars, we need to consider only the co-rotating orbit, so from here and thereafter we take into account only the plus sign in Eq.~(\ref{eight}) and we write $\Omega^{+}_{\rm orb}(r)=\Omega_{\rm orb}(r)$.

For the Hartle external solution given by Eq.~(\ref{ht1}) we obtain

\begin{equation}\label{eq:omegaKep}
\Omega_{K}^{J\neq0}(r)=\sqrt{\frac{M}{r^3}}\left[1- j F_{1}(r)+j^2F_{2}(r)+q F_{3}(r)\right] \;,
\end{equation}
where $j=J/M^2$ and $q=Q/M^3$ are the dimensionless angular momentum and quadrupole moment. The analytical expressions of the functions $F_i$ are given by
\begin{align*}
F_1&=\left(\frac{M}{r}\right)^{3/2} \;,\\
F_2&=\frac{48M^7-80M^6r+4M^5r^2-18M^4r^3}{16M^2r^4(r-2M)}+\frac{40M^3r^4+10M^2r^5+15Mr^6-15r^7}{16M^2r^4(r-2M)}+F\;,\\
F_3&=\frac{6M^4-8M^3r-2M^2r^2-3Mr^3+3r^4}{16M^2r(r-2M)/5}-F \;,\\
F&=\frac{15(r^3-2M^3)}{32M^3}\ln\frac{r}{r-2M} \;.\\
\end{align*}

The maximum angular velocity for a rotating star at the mass-shedding limit is the Keplerian angular velocity evaluated at the equator ($r=R_{\rm eq}$), i.e.
\begin{equation}\label{eq:omegaK1}
\Omega_K^{J\neq0}=\Omega_{\rm orb}(r=R_{\rm eq})\;.
\end{equation}
In the static case i.e. when $j=0$ hence $q=0$ and $\delta M=0$ we have the well-known Schwarzschild solution and the orbital angular velocity for a test particle $\Omega_K^{J=0}$ on the surface ($r=R$) of the neutron star is given by
\begin{equation}\label{eq:omegaK2}
\Omega_K^{J=0}=\sqrt{\frac{M^{J=0}}{R_{M^{J=0}}^3}}\;.
\end{equation}
\subsection{Gravitational binding energy}\label{subsec:3.3}

Besides the above stability requirements, one should check if the neutron star is gravitationally bound. In the non-rotating case, the binding energy of the star can be computed as
\begin{equation}\label{eq:W0}
W_{J=0}=M_0-M^0_{\rm rest}\;,\qquad M^0_{\rm rest}=m_b A_{J=0}\;,
\end{equation}
where $M^0_{\rm rest}$ is the rest-mass of the star, $m_b$ is the rest-mass per baryon, and $A_{J=0}$ is the total number of baryons inside the star. So the non-rotating star is considered bound if $W_{J=0}<0$.

In the slow rotation approximation the total binding energy is given by Ref.~\cite{1968ApJ...153..807H}
\begin{equation}\label{eq:Wrot}
W_{J\neq 0}=W_{J=0}+\delta W\;,\qquad  \delta W=\frac{J^2}{R^3}-\int_{0}^R 4\pi r^2 B(r)dr\;,
\end{equation}
where
\begin{align}
B(r)&=({\cal E}+P)p^*_0\bigg\{ \frac{d{\cal E}}{dP}\left[\left(1-\frac{2 M}{r}\right)^{-1/2}-1\right]- \frac{d u}{dP}\left(1-\frac{2 M}{r}\right)^{-1/2}\bigg\}\nonumber \\
&+({\cal E}-u)\left(1-\frac{2M}{r}\right)^{-3/2}\bigg[\frac{m_0}{r}+\frac{1}{3}j^2r^2\bar{\omega}^2\bigg]\nonumber \\&-\frac{1}{4\pi r^2}\left[ \frac{1}{12}j^2 r^4 \left(\frac{d\bar{\omega}}{dr}\right)^2-\frac{1}{3}\frac{dj^2}{dr}r^3 \bar{\omega}^2 \right]\;,
\end{align}
where $u={\cal E}-m_b n_b$ is the internal energy of the star, with $n_b$ the baryon number density.

\section{Structure of uniformly rotating neutron stars}\label{sec:4}

We show now the results of the integration of the Hartle equations for the globally and locally charge neutrality neutron stars; see e.g.~Fig.~\ref{fig:Model}. Following Refs.~\cite{2012NuPhA.883....1B, belvedere2014}, we adopt, as an example, globally neutral neutron stars with a density at the edge of the crust equal to the neutron drip density, ${\it \rho}_{\rm crust}={\it \rho}_{\rm drip}\approx 4.3\times 10^{11}$ g cm$^{-3}$.

\subsection{Secular instability boundary}\label{sec:4.1}

\begin{figure}[!hbtp]
\centering
\includegraphics[width=0.75\hsize,clip]{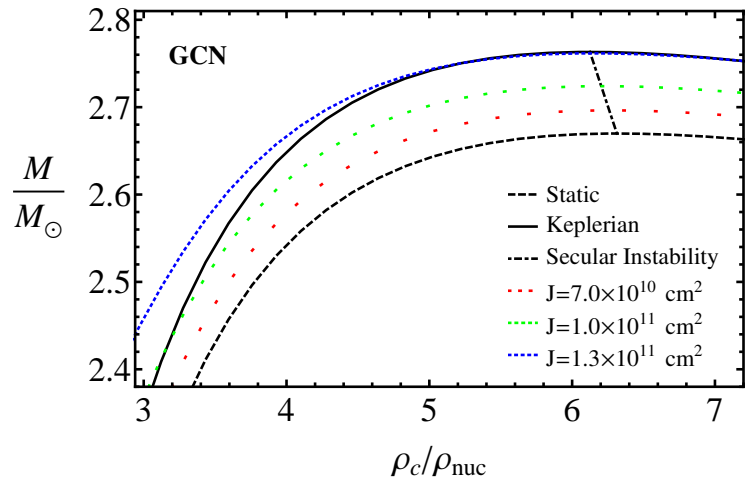}
\caption{Total mass is shown as a function of central density for neutron stars with global charge neutrality. The mass is given in solar mass $M_{\odot}=1.98\times10^{33}$g  and the density is normalized to nuclear density ${\it \rho}_{\rm nuc}=2.7\times10^{14}$g cm$^{-3}$. The solid curve represents the configuration at the Keplerian mass-shedding sequence, the dashed curve represents the static sequence, the dotted curves represent the $J$=constant sequences. The doted-dashed line joins all the turning points of the $J$=constant sequences, so it determines the axisymmetric secular instability boundary.}\label{fig:MtotvsRhoG}
\end{figure}

In Fig.~\ref{fig:MtotvsRhoG} we show the mass-central density curve for globally neutral neutron stars in the region close to the axisymmetric stability boundaries. Particularly, we construct some $J$=constant sequences to show that indeed along each of these curves there exist a maximum mass (turning point). The line joining all the turning points determines the secular instability limit. In Fig.~\ref{fig:MtotvsRhoG} the axisymmetric stable zone is on the left side of the instability line.

\begin{figure}[!hbtp]
\centering
\includegraphics[width=0.75\hsize,clip]{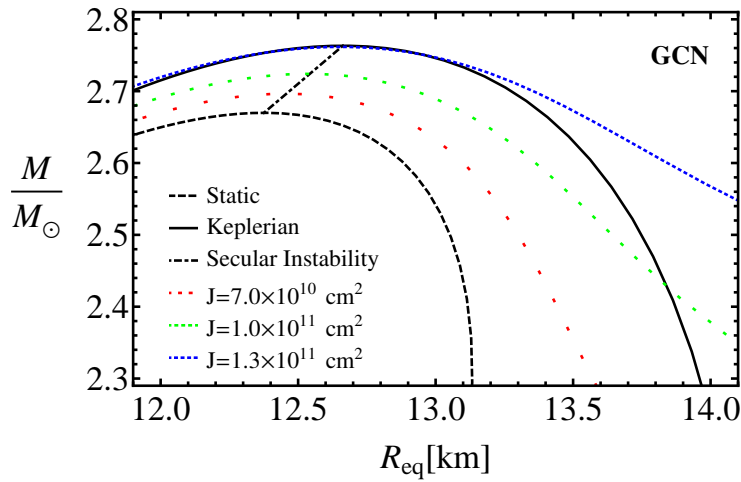}
\caption{Total mass is shown as a function of equatorial radius for neutron stars with global charge neutrality. The solid curve represents the configuration at Keplerian mass-shedding sequence, the dashed curve represents the static sequence, the dotted curves represent the $J$-constant sequences. The doted-dashed line is the  secular instability boundary.}\label{fig:MtotvsReqG}
\end{figure}

Clearly we can transform the mass-central density relation in a mass-radius relation. In Fig.~\ref{fig:MtotvsReqG} we show the mass versus the equatorial radius of the neutron star that correspond to the range of densities of Fig.~\ref{fig:MtotvsRhoG}. In this plot the stable zone is on the right side of the instability line.

We can construct a fitting curve joining the turning points of the $J$=constant sequences line which determines the axisymmetric secular instability boundary. Defining $M_{{\rm max},0}$ as the maximum stable mass of the non-rotating neutron star constructed with the same EOS, we find that for globally neutral configurations the instability line is well fitted by the function
\begin{align}\label{eq:SecularG}
\frac{M^{\rm GCN}_{\rm sec}}{M_\odot}&=21.22-6.68\frac{M_{{\rm max},0}^{\rm GCN}}{M_\odot}-\left(77.42-28\frac{M_{{\rm max},0}^{\rm GCN}}{M_\odot}\right)\left(\frac{R_{\rm eq}}{10\,{\rm km}}\right)^{-6.08}\;,
\end{align}
where $12.38\,{\rm km}\lesssim R_{\rm eq}\lesssim 12.66\,{\rm km}$, and $M_{{\rm max},0}^{\rm GCN}\approx 2.67 M_\odot$.

\begin{figure}[!hbtp]
\centering
\includegraphics[width=0.75\hsize,clip]{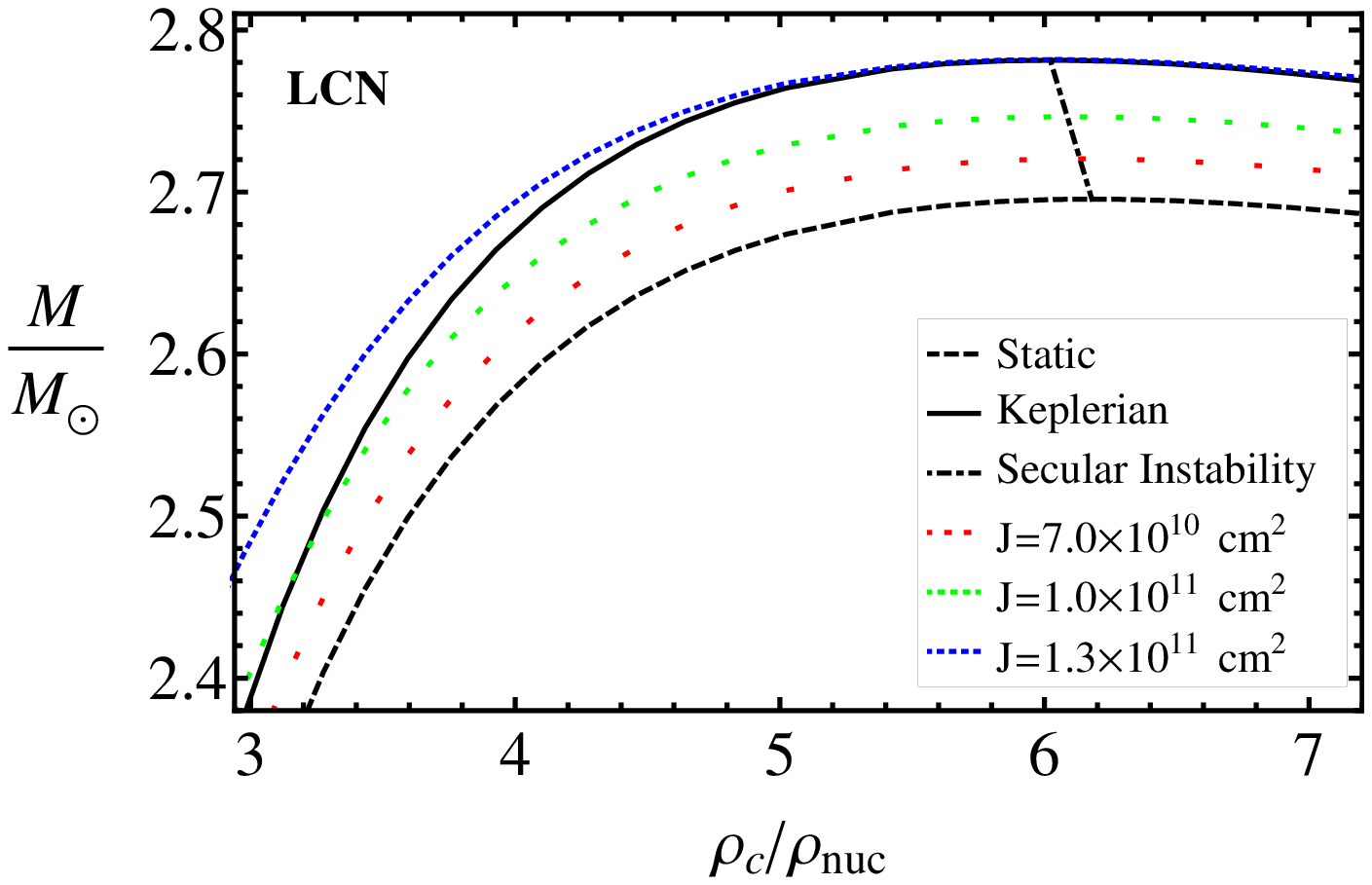}
\caption{Total mass is shown as a function of central density for neutron stars with local charge neutrality. The mass is given in solar mass $M_{\odot}=1.98\times10^{33}$g  and the density is normalized to nuclear density ${\it \rho}_{\rm nuc}=2.7\times10^{14}$g cm$^{-3}$.  The solid curve represents the configuration at Keplerian mass-shedding sequence, the dashed curve represents the static sequence, the dotted curves represent the $J$=constant sequences. The doted-dashed line determines the axisymmetric secular instability boundary.}\label{fig:MtotvsRhoL}
\end{figure}

\begin{figure}[!hbtp]
\centering
\includegraphics[width=0.75\hsize,clip]{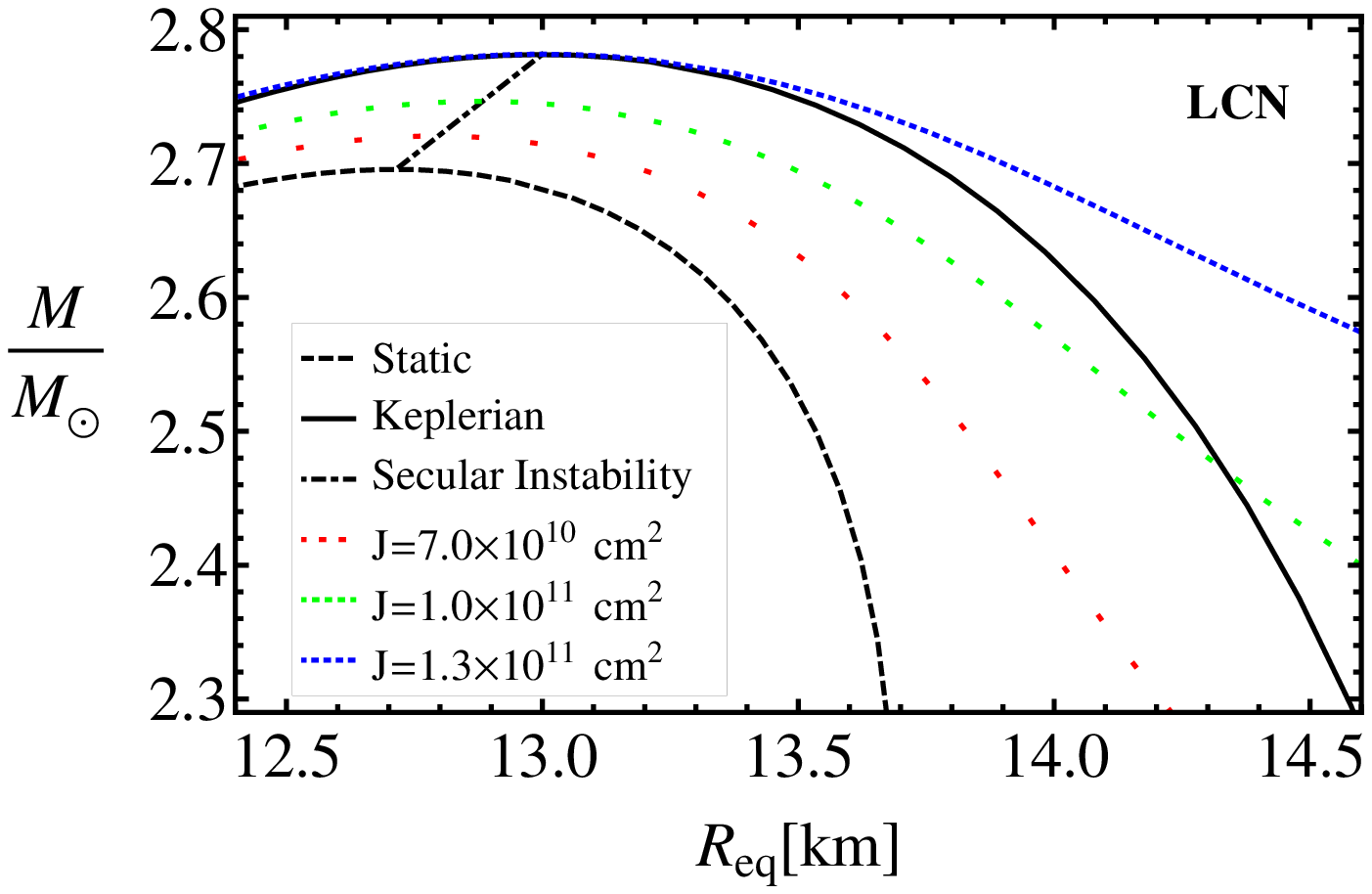}
\caption{Total mass is shown as a function of equatorial radius for neutron stars with local charge neutrality. The solid curve represents the configuration at Keplerian mass-shedding sequence, the dashed curve represents the static sequence, the dotted curves represent the $J$=constant sequences. The doted-dashed line is the axisymmetric secular instability boundary.}\label{fig:MtotvsReqL}
\end{figure}

The turning points of locally neutral configurations in the mass-central density plane are shown in Fig.~\ref{fig:MtotvsRhoL}. the corresponding mass-equatorial radius plane is plotted in Fig.~\ref{fig:MtotvsReqL}.

For locally neutral neutron stars, the secular instability line is fitted by
\begin{align}\label{eq:SecularL}
\frac{M^{\rm LCN}_{\rm sec}}{M_\odot}&=20.51-6.35\frac{M_{{\rm max},0}^{\rm LCN}}{M_\odot}-\left(80.98-29.02\frac{M_{{\rm max},0}^{\rm LCN}}{M_\odot}\right)\left(\frac{R_{\rm eq}}{10\,{\rm km}}\right)^{-5.71}\;,
\end{align}
where $12.71\,{\rm km}\lesssim R_{\rm eq}\lesssim 13.06\,{\rm km}$, and $M_{\rm max,0}^{\rm LCN}\approx 2.70 M_\odot$.

\subsection{Keplerian mass-shedding sequence}\label{sec:4.2}

We turn now to analyze in detail the behavior of the different properties of the neutron star along the Keplerian mass-shedding sequence. 

\subsubsection{Maximum mass and rotation frequency}\label{sec:4.2.1}

\begin{figure}[!hbtp]
\centering
\includegraphics[width=0.75\hsize]{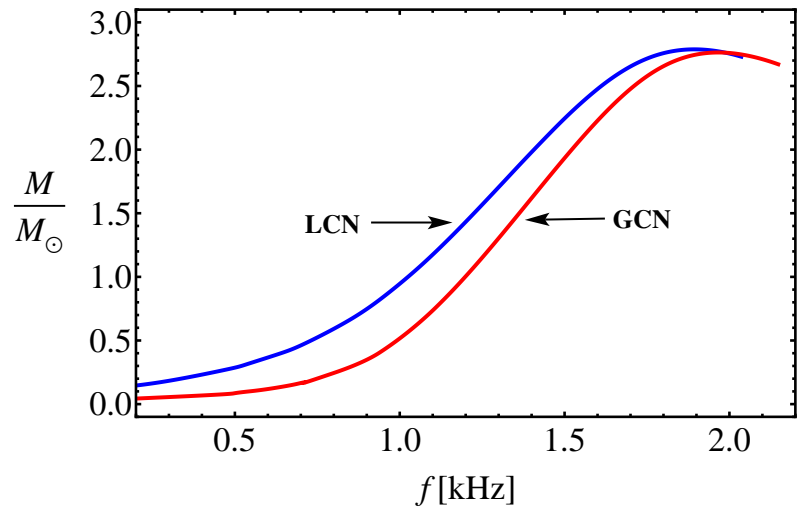}
\caption{Total mass is shown as a function of rotational Keplerian frequency both for the global (red) and local (blue) charge neutrality cases.}\label{fig:Mtotvsf}
\end{figure}

The total mass of the rotating star is computed from Eq.~(\ref{eq:Mrot}). In Fig.~\ref{fig:Mtotvsf}  the total mass of the neutron star is shown as a function of the rotation frequency for the Keplerian sequence. It is clear that for a given mass, the rotational frequency is higher for a globally neutral neutron star with respect to the locally neutral one.

\begin{figure}[!hbtp]
\centering
\includegraphics[width=0.75\hsize]{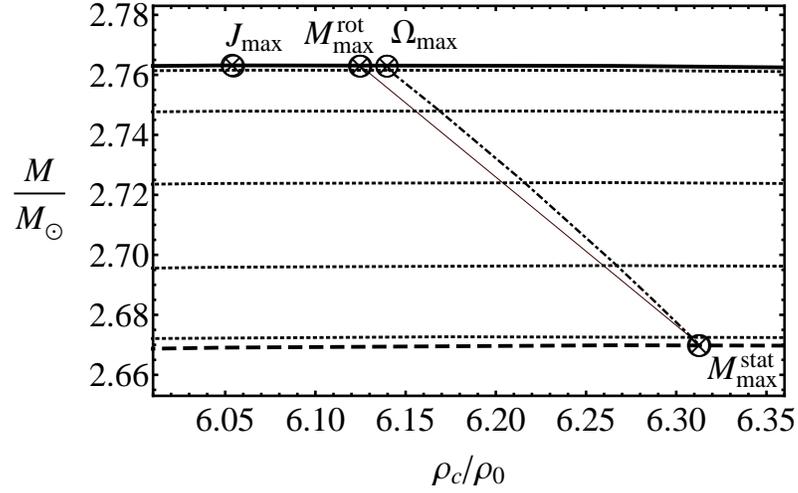}
\caption{Total mass versus central density for the global charge neutrality case. The dashed curve is the static configurations, the solid curve is the Keplerian mass-shedding configurations, the dotted curves are the $J$=constant sequences, the dotted-dashed line is axisymmetric secular instability boundary. The thin line  joins $M_{max}^{rot}$ with $M_{max}^{stat}$.}\label{fig:Mmaxrho}
\end{figure}

The configuration of maximum mass, $M^{J\neq 0}_{\rm max}$, is obtained along the Keplerian sequence, and it is found before the secular instability line crosses the Keplerian curve. Thus, the maximum mass configuration is secularly stable. This implies that the configuration with maximum rotation frequency, $f_{\rm max}$, is located beyond the maximum mass point, specifically at the crossing point between the secular instability and the Keplerian mass-shedding sequence. The results are summarized in Table~\ref{tab:MRfmax} and shown in Fig.~\ref{fig:Mmaxrho}.

In Fig.~\ref{fig:Mmaxrho} details near the maximum masses are illustrated. Here we focus on the definition of maximum rotating mass $M_{max}^{rot}$, maximum static mass, maximum angular momentum $J_{max}$ and maximum angular velocity (minimum rotation period) $\Omega_{max} (P_{min})$. Note, that $\Omega_{max}$ is determined along the turning points of $J$=constant sequences (axisymmetric secular instability line) what is consistent with the results of Stergioulas and Friedman (1995) \cite{sterfried1995}. At large scales the difference between axisymmetric secular instability line and the line joining $M_{max}^{rot}$ with $M_{max}^{stat}$ can not be seen (for more details see \cite{haenselbook}).

It is important to discuss briefly the validity of the present perturbative solution for the computation of the properties of maximally rotating neutron stars. The expansion of the radial coordinate of a rotating configuration $r(R,\theta)$ in powers of angular velocity is written as \cite{1967ApJ...150.1005H}
\begin{equation}
r \approx R+\xi(R,\theta)+O(\Omega^4) \; ,
\end{equation}
where $\xi$ is the difference in the radial coordinate, $r$, between a point located at the polar angle $\theta$ on the surface of constant density ${\it \rho}(R)$ in the rotating configuration, and the point located at the same polar angle on the same constant density surface in the static configuration.
In the slow rotation regime, the fractional displacement of the surfaces of constant density due to the rotation have to be small, namely $\xi(R,\theta)/R\ll 1$, where $\xi(R,\theta)=\xi_0(R)+\xi_2(R)P_2(\cos\theta)$ and $\xi_0(R)$ and $\xi_2(R)$ are function of $R$, proportional to $\Omega^2$. From Table~\ref{tab:MRfmax}, we can see that the configuration with the maximum possible rotation frequency has a maximum fractional displacement $\delta R_{\rm eq}^{\rm max}=\xi(R,\pi/2)/R$ as low as $\approx 2\%$ and $\approx 3\%$, for the globally and locally neutral neutron stars respectively.

In this line, it is worth quoting the results of Ref.~\cite{benhar05}, where it has been shown that the inclusion of a third-order expansion $\Omega^3$ in the Hartle's method improves the value of the maximum rotation frequency by less than 1\% for different EOS. The reason for this is that as mentioned above, along the Keplerian sequence the deviations from sphericity decrease with density 
(see Fig.~\ref{fig:eccvsrho}), which ensures the accuracy of the perturbative solution.

Turning to the increase of the maximum mass, it has been shown in Ref.~\cite{1992ApJ...390..541W} that the mass of maximally rotating neutron stars, computed with the Hartle's second order approximation, is accurate within an error as low as $\lesssim 4\%$.

\begin{table}
\caption{Maximum static mass $M^{J=0}_{\rm max}$ and corresponding static radius $R^{J=0}_{\rm max}$ of neutron stars as computed in Ref.~\cite{2012NuPhA.883....1B};  maximum rotating mass $M^{J\neq0}_{\rm max}$ and corresponding equatorial radius $R^{J\neq0}_{\rm max}$ of neutron stars as given in Refs.~\cite{Boshkayev2012thesis, belvedere2014}; increase in mass $\delta M_{\rm max}$ and radius $\delta R_{\rm eq}^{\rm max}$ of the maximum mass configuration with respect to its non-rotating counterpart; maximum rotation frequency $f_{\rm max}$ and corresponding minimum period $P_{\rm min}$.}
\centering
\begin{tabular}{c c c}
& Global Neutrality & Local Neutrality \\
\hline
$M^{J=0}_{\rm max}$ $(M_\odot)$ & 2.67 & 2.70\\
$R^{J=0}_{\rm max}$ (km) & 12.38 & 12.71\\
$M^{J\neq 0}_{\rm max}$ $(M_\odot)$ & 2.76 & 2.79\\
$R^{J\neq 0}_{\rm max}$ (km) & 12.66 & 13.06\\
$\delta M_{\rm max}$ & 3.37\% & 3.33\%\\
$\delta R_{\rm eq}^{\rm max}$ & 2.26\% & 2.75\%\\
$f_{\rm max}$ (kHz)& 1.97 & 1.89\\
$P_{\rm min}$ (ms)& 0.51 & 0.53\\
\hline
\end{tabular}
\label{tab:MRfmax}
\end{table}

We compute the gravitational binding energy of the neutron star from Eq.~(\ref{eq:Wrot}) as a function of the central density and angular velocity. We make this for central densities higher than the nuclear density, thus we impose the neutron star to have a supranuclear hadronic core. 

\section{Neutron star mass-radius relation}\label{sec:5}
%
\begin{figure}[!hbtp]
\centering
\includegraphics[width=0.75\hsize]{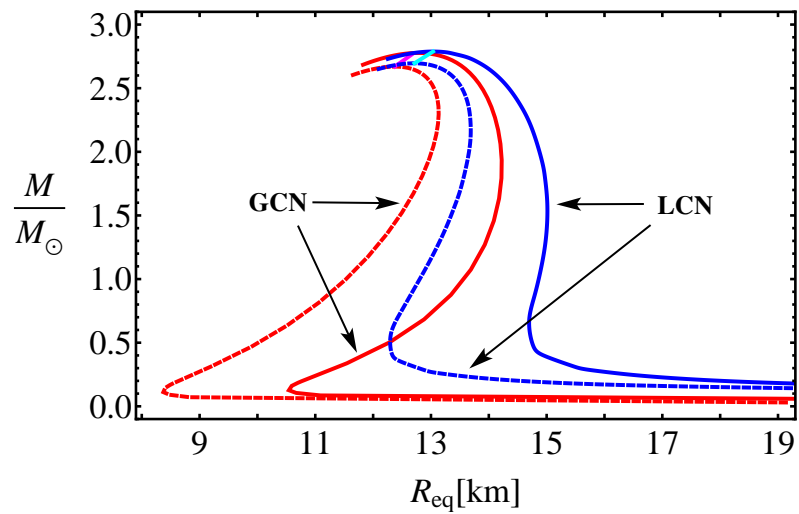}
\caption{Total mass versus total equatorial radius for the global (red) and local (blue) charge neutrality cases. The dashed curves represent the static configurations, while the solid lines are the uniformly rotating neutron stars. The pink-red and light-blue color lines define the secular instability boundary for the globally and locally neutral cases, namely the lines given by Eqs.~(\ref{eq:SecularG}) and (\ref{eq:SecularL}), respectively.}\label{fig:MtotvsRtot}
\end{figure}

We summarize now the above results in form of a new mass-radius relation of uniformly rotating neutron stars, including the Keplerian and secular instability boundary limits. In Fig.~\ref{fig:MtotvsRtot} we show a summary plot of the equilibrium configurations of rotating neutron stars. In particular we show the total mass versus the equatorial radius: the dashed lines represent the static (non-rotating, $J=0$) sequences, while the solid lines represent the corresponding Keplerian mass-shedding sequences. The secular instability boundaries are plotted in pink-red and light blue color for the global and local charge neutrality cases, respectively.

It can be seen that due to the deformation for a given mass the radius of the rotating case is larger than the static one, and similarly the mass of the rotating star is larger than the corresponding static one. It can be also seen that the configurations obeying global charge neutrality are more compact with respect to the ones satisfying local charge neutrality.

In general, the region enclosed by the static, Keplerian mass-shedding and secular instability sequences is termed as the stability region for uniformly rotating neutron stars. All stable configurations are inside this region. From a practical point of view it is important to construct the stability region as it allows one to do simple and at the same interesting science without invoking sophisticated numerical simulations, which require powerful super computers and intricate techniques. For instance, for a given value of the angular velocity (rotation period) one can construct the $\Omega$=constant sequence. This sequence intersects the stability region at two points and determines the maximum and minimum values of all physical quantities, describing the structure of rotating neutron stars such as mass, radius, moment of inertia, angular momentum, quadrupole moment etc. Thus this procedure  allows one to estimate the range of all the quantities for $\Omega$=constant sequence. 

The same procedure is valid if there is a necessity to construct the constant baryon (rest) mass sequence. In analogy to the previous case, the constant rest mass sequence also intersects the stability region at two points. Here again one can estimate the upper and lower bounds of all the quantities corresponding to the constant rest mass sequence. The so-called spin-up and spin-down effects emerges as a results of this sequence. Of course, there are other crucial applications of the stability region related to the post merge epoch of neutron stars, binary systems, gravitational wave, physics of gamma ray burst etc. For more details see Ref.~\cite{belvedere2014jkps} and references therein.   

\section{Moment of inertia}\label{sec:6}
\begin{figure}[!hbtp]
\centering
\includegraphics[width=0.75\hsize,clip]{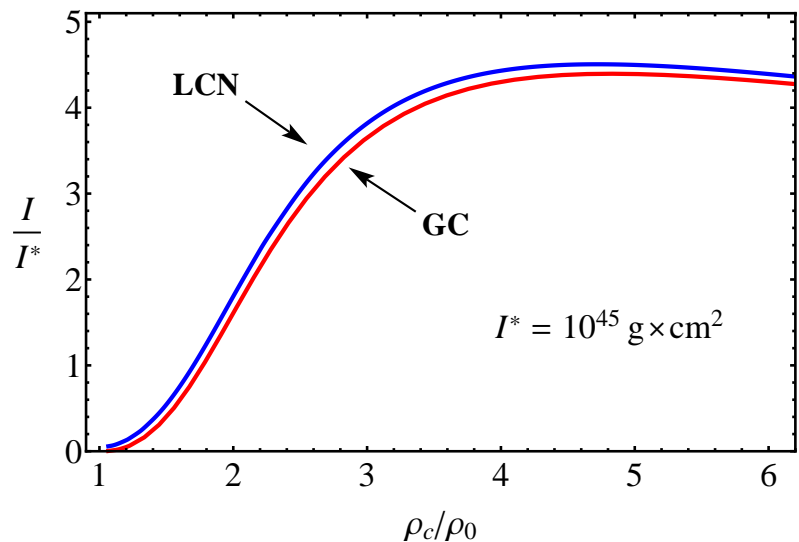}
\caption{Total moment of inertia as a function central density for globally (red) and locally (blue) neutral non-rotating neutron stars.}\label{fig:inertiarho}
\end{figure}

The moment of inertia $I$ of relativistic stars can be computed from the relation
\begin{equation}\label{eq:I}
I = \frac{J}{\Omega}\;,
\end{equation}
where $J$ is the angular momentum and $\Omega$ is angular velocity, as before, are related via Eq.~(\ref{eq:Jomega}). Since $J$ is a first-order quantity and so proportional to $\Omega$, the moment of inertia given by Eq.~(\ref{eq:I}) does not depend on the angular velocity and does not take into account deviations from the spherical symmetry. This implies that Eq.~(\ref{eq:J}) gives the moment of inertia of the non-rotating unperturbed seed object. In order to find the perturbation to $I$, say $\delta I$, the perturbative treatment has to be extended to the next order $\Omega^3$, in such a way that $I=I_0+\delta I =(J_0+\delta J)/\Omega$, becomes of order $\Omega^2$, with $\delta J$ of order $\Omega^3$ \cite{1973Ap&SS..24..385H,benhar05}. In this chapter we keep the solution up to second order and therefore we proceed to analyze the behavior of the moment of inertia for the static configurations. In any case, even the fastest observed pulsars rotate at frequencies much lower than the Keplerian rate, and under such conditions we expect that the moment of inertia can be approximated with high accuracy by the one of the corresponding static configurations.

In Fig.~\ref{fig:inertiarho} we show the behavior of the total moment of inertia, i.e. $I=I_{\rm core}+I_{\rm crust}$, with respect to the central density for both globally and locally neutral static neutron stars.

We can see from Fig.~\ref{fig:inertiarho} that the total moment of inertia is quite similar for both global and local charge neutrality cases. This is due to the fact that the globally neutral configurations differ from the locally ones mostly in the structure of the crust, which however contributes much less than the neutron star core to the total moment of inertia.

In order to study the single contribution of the core and the crust to the moment of inertia of the neutron star, we shall use the integral expression for the moment of inertia. Multiplying Eq.~(\ref{eq:baromega}) by $r^3$ and taking the integral of it we obtain
\begin{align}
I(r)&=-\frac{2}{3}\int_0^r r^3 \frac{d j}{dr}\frac{\bar{\omega}(r)}{\Omega}dr=\frac{8\pi}{3}\int_0^r r^4 ({\cal E}+P) e^{(\lambda-\nu)/2}\frac{\bar{\omega}(r)}{\Omega}dr\;,
\end{align}
where the integration is carried out in the region of interest. Thus, the contribution of the core, $I_{\rm core}$, is obtained integrating from the origin up to the radius of the core, and the contribution of the crust, $I_{\rm crust}$, integrating from the base of the crust to the total radius of the neutron star.

\section{Deformation of the neutron star}\label{sec:7}

In this section we explore the deformation properties of the neutron star. The behavior of the eccentricity, the rotational to gravitational energy ratio, as well as the quadrupole moment, are investigated as a function of the central density and rotation frequency of the neutron star.
%
\begin{figure}[!hbtp]
\centering
\includegraphics[width=0.75\hsize,clip]{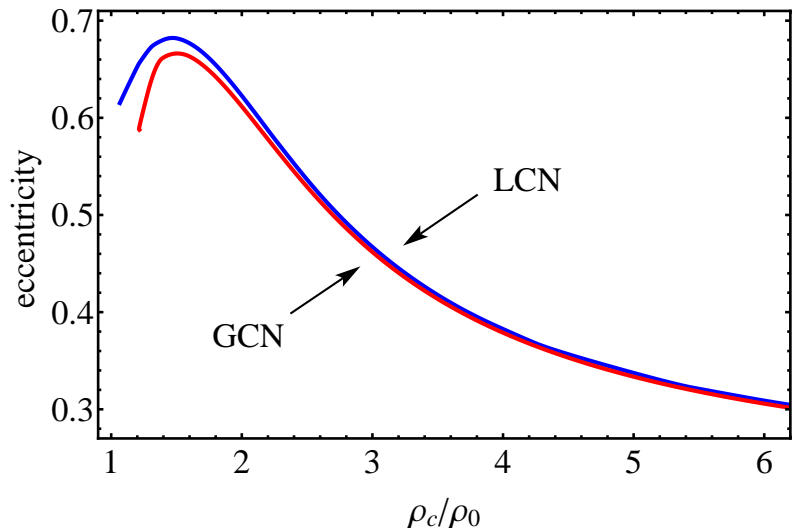}
\caption{Eccentricity (\ref{eq:eccentricity}) as a function of central density for the Keplerian sequence in both global (red) and local (blue) charge neutrality cases.}\label{fig:eccvsrho}
\end{figure}

\subsection{Eccentricity}\label{sec:4.5.1}

An indicator of the neutron star deformation degree can be estimated with its eccentricity
\begin{equation}\label{eq:eccentricity}
\epsilon=\sqrt{1-\left(\frac{R_p}{R_{\rm eq}}\right)^2}\;,
\end{equation}
where $R_p$ and $R_{\rm eq}$ are the polar and equatorial radii of the rotating deformed configuration. Thus, $\epsilon=0$ determines the spherical limit and $0<\epsilon<1$ corresponds to oblate configurations.

We can see in Fig.~\ref{fig:eccvsrho} that for larger central densities the neutron stars decrease their oblateness and the configurations tend to be a more spherical.

\begin{figure*}[!hbtp]
\centering
\includegraphics[width=0.48\columnwidth]{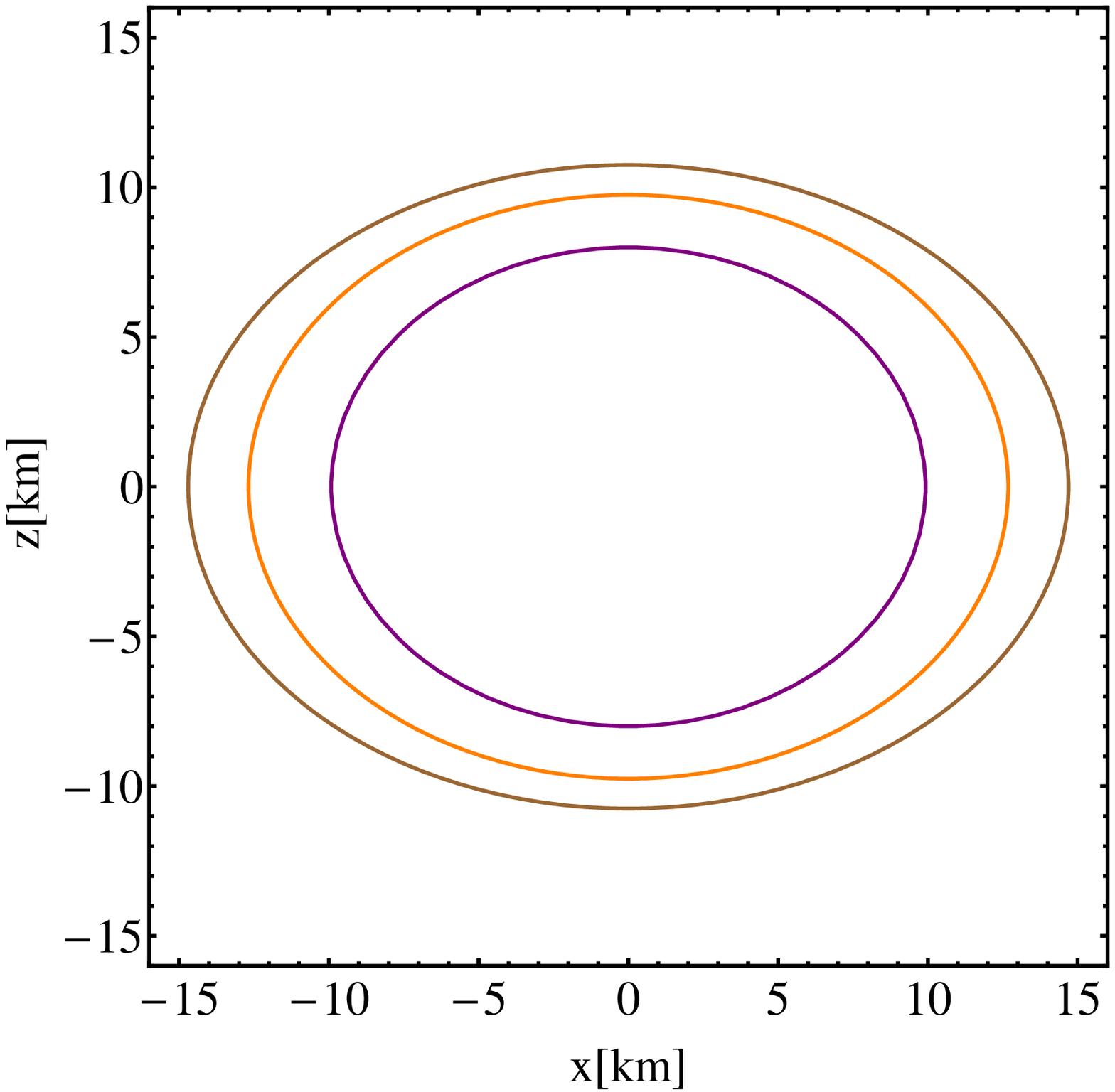} \quad \includegraphics[width=0.48\columnwidth]{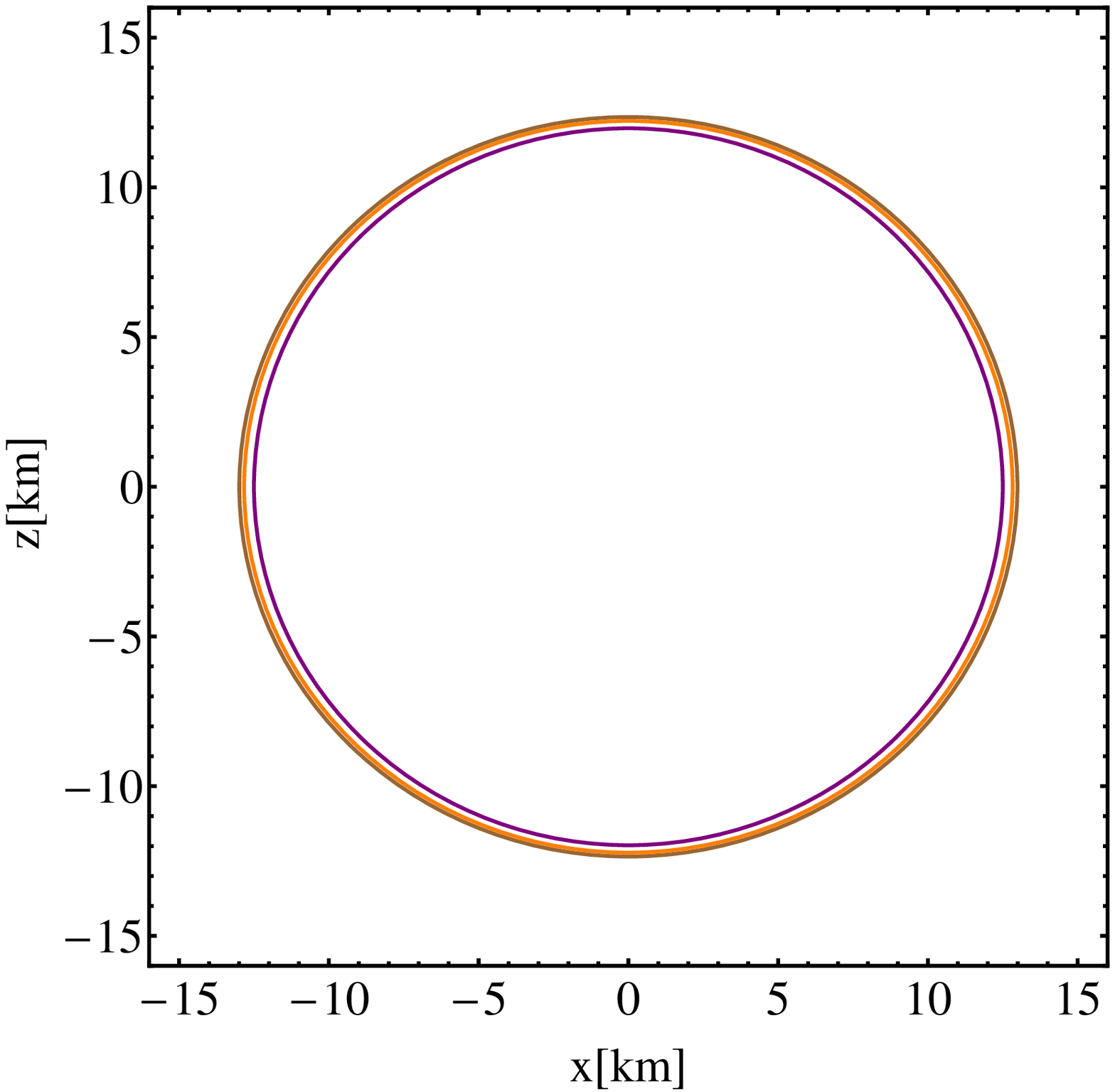}
\caption{Cross section in the plane passing through the rotation axis of a neutron star with the traditional locally neutral TOV treatment at frequency $f=852$Hz and $f=1900$Hz. The total rotating gravitational mass is $M=0.67 M_\odot$ and $M=2.78 M_\odot$, the central density ${\it \rho}_c=1.445{\it \rho}_{\rm nuc}$ and ${\it \rho}_c=6{\it \rho}_{\rm nuc}$ (left and right panels). The contours are the lines of constant density. The inner contour is the core-inner crust interface, corresponding to the nuclear density; the outer contour is the stellar surface, and the middle (intermediate) contour corresponds to the neutron drip density corresponding to the inner-outer crusts interface.}\label{fig:lcn_shape}
\end{figure*}
\begin{figure*}[!hbtp]
\centering
\includegraphics[width=0.48\columnwidth]{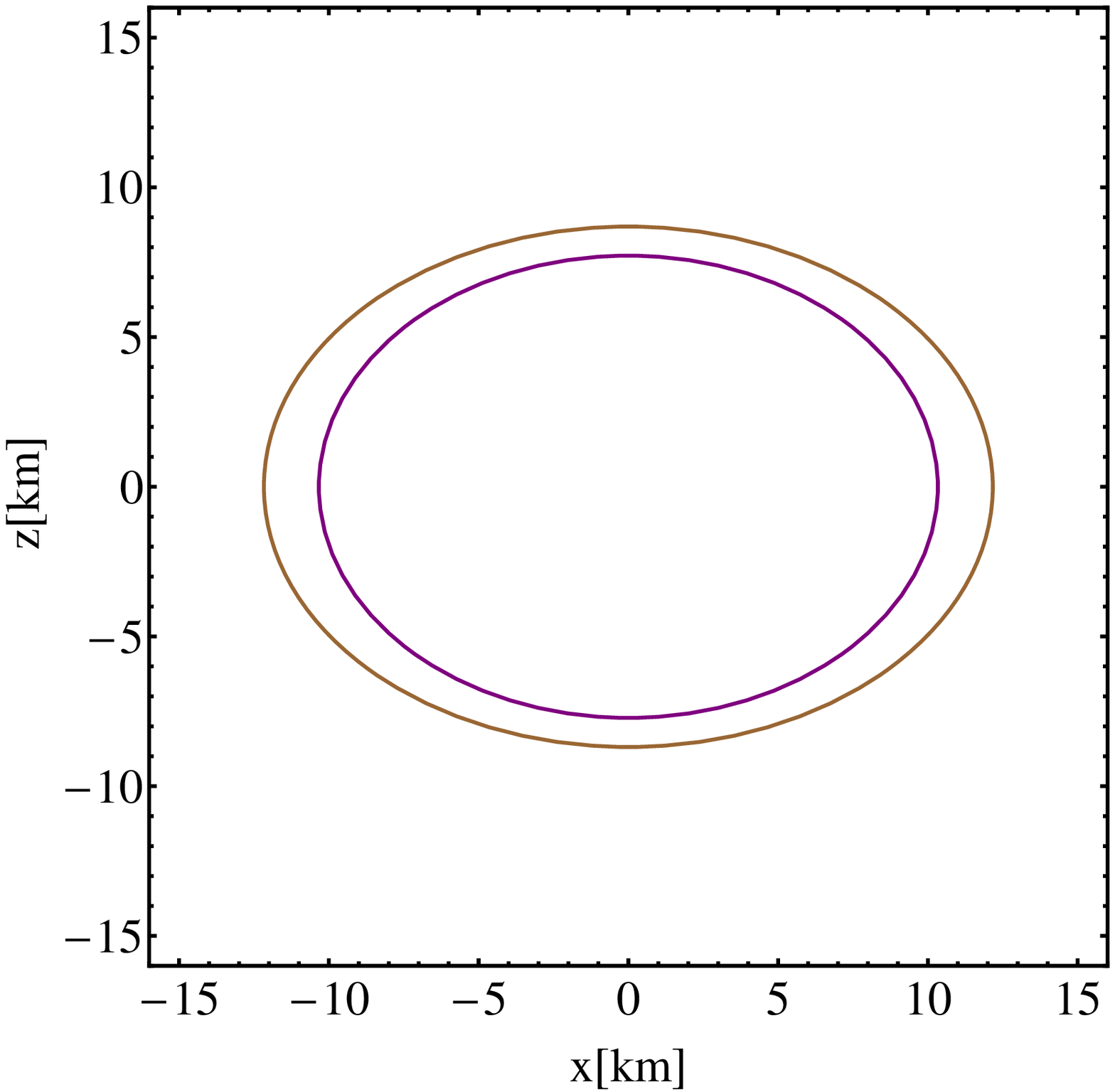}\quad \includegraphics[width=0.48\columnwidth]{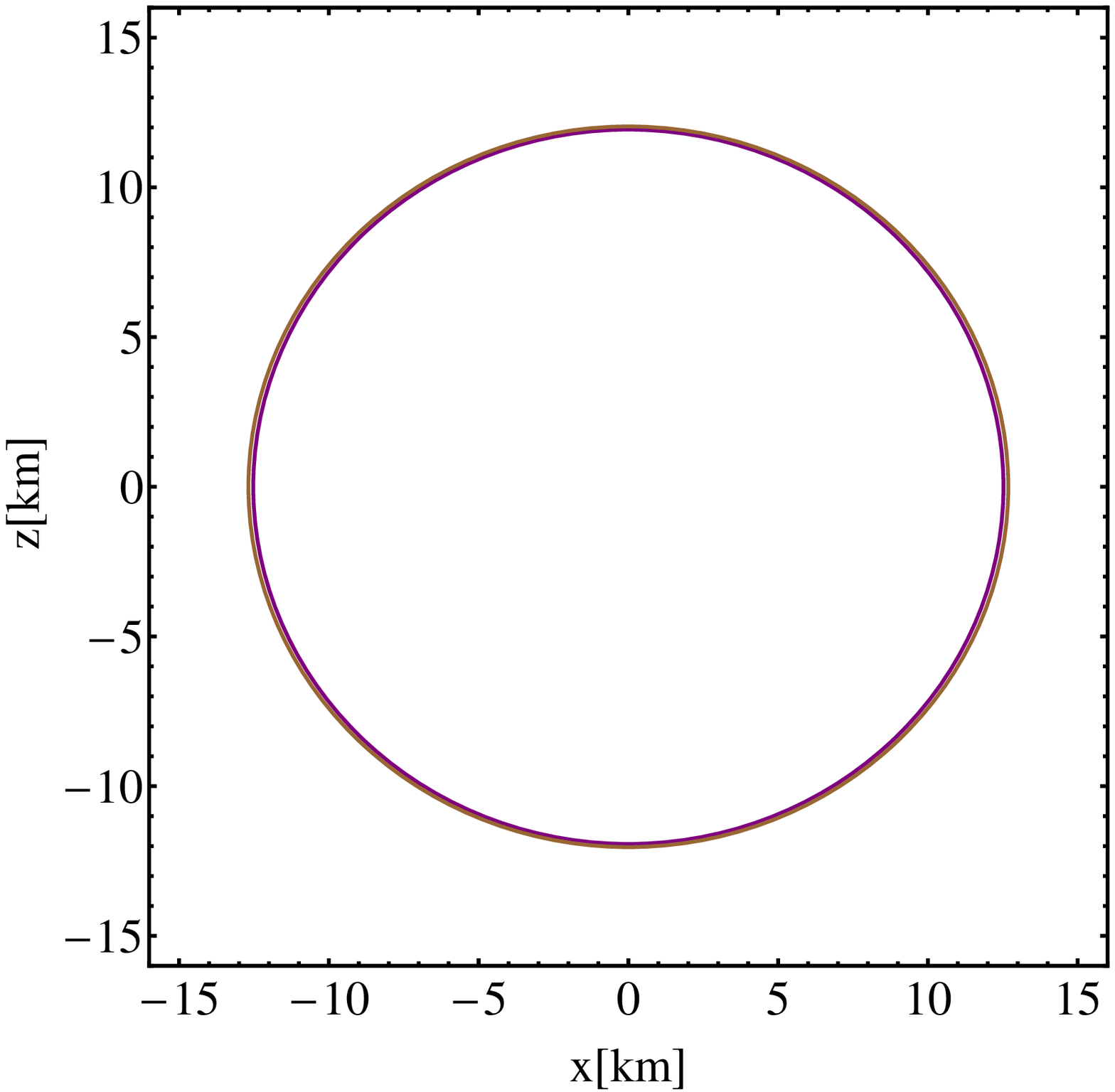}
\caption{Cross section in the plane passing through the rotation axis of a neutron star with the new globally neutral equilibrium configurations presented in Refs. \cite{Boshkayev2012thesis} at frequency $f=986$Hz and $f=1942$Hz. The total rotating gravitational mass is $M=0.488M_\odot$ and $M=2.763M_\odot$, the central density ${\it \rho}_c=1.445{\it \rho}_{\rm nuc}$ and ${\it \rho}_c=6{\it \rho}_{\rm nuc}$ (left and right panels). The contours are the lines of the constant density. The inner  contour is the core-crust interface (corresponds to the nuclear and neutron drip densities); the outer contour is the stellar surface.}\label{fig:gcn_shape}
\end{figure*}

The shape of rotating neutron stars becomes less oblate with the increasing central density (see Figs.~\ref{fig:lcn_shape},~\ref{fig:gcn_shape} for details). The size of the core initially increases then after reaching its maximum decreases with the increasing central density. The thickness of the crusts in both global and local neutrality cases gradually decreases with the increasing central density. However the radii of the crusts behave similarly to the radius of the core.  Close to the maximum rotating mass, even though the configurations rotate at the Keplerian rate, the shape becomes almost spherical, but still oblate (see the right panels of Figs.~\ref{fig:lcn_shape},~\ref{fig:gcn_shape} for details). Thus, the system become more gravitationally bound.

\begin{figure*}[!hbtp]
\centering
\includegraphics[width=0.75\columnwidth]{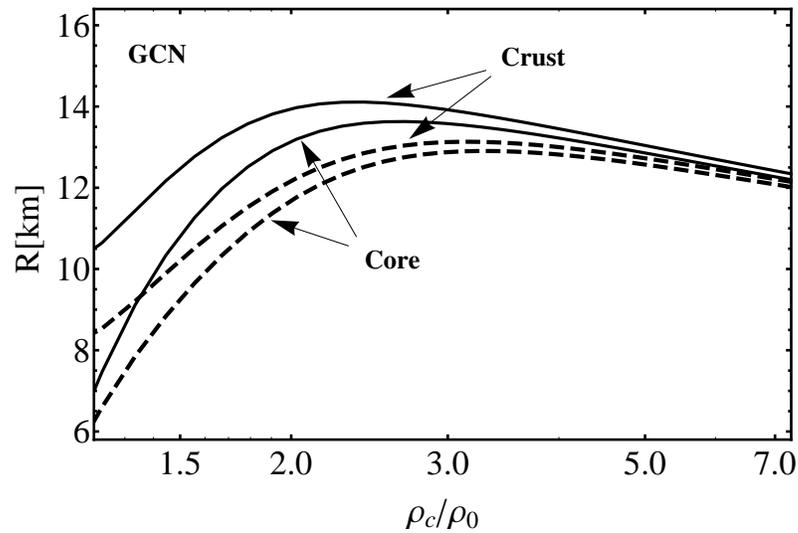}
\caption{Radius versus central density for globally neutral configuration. All dashed curves are for the static configurations. All solid curves are the equatorial radii for the Keplerian sequence. }\label{fig:gcn_Rrho}
\end{figure*}

\begin{figure*}[!hbtp]
\centering
 \includegraphics[width=0.75\columnwidth]{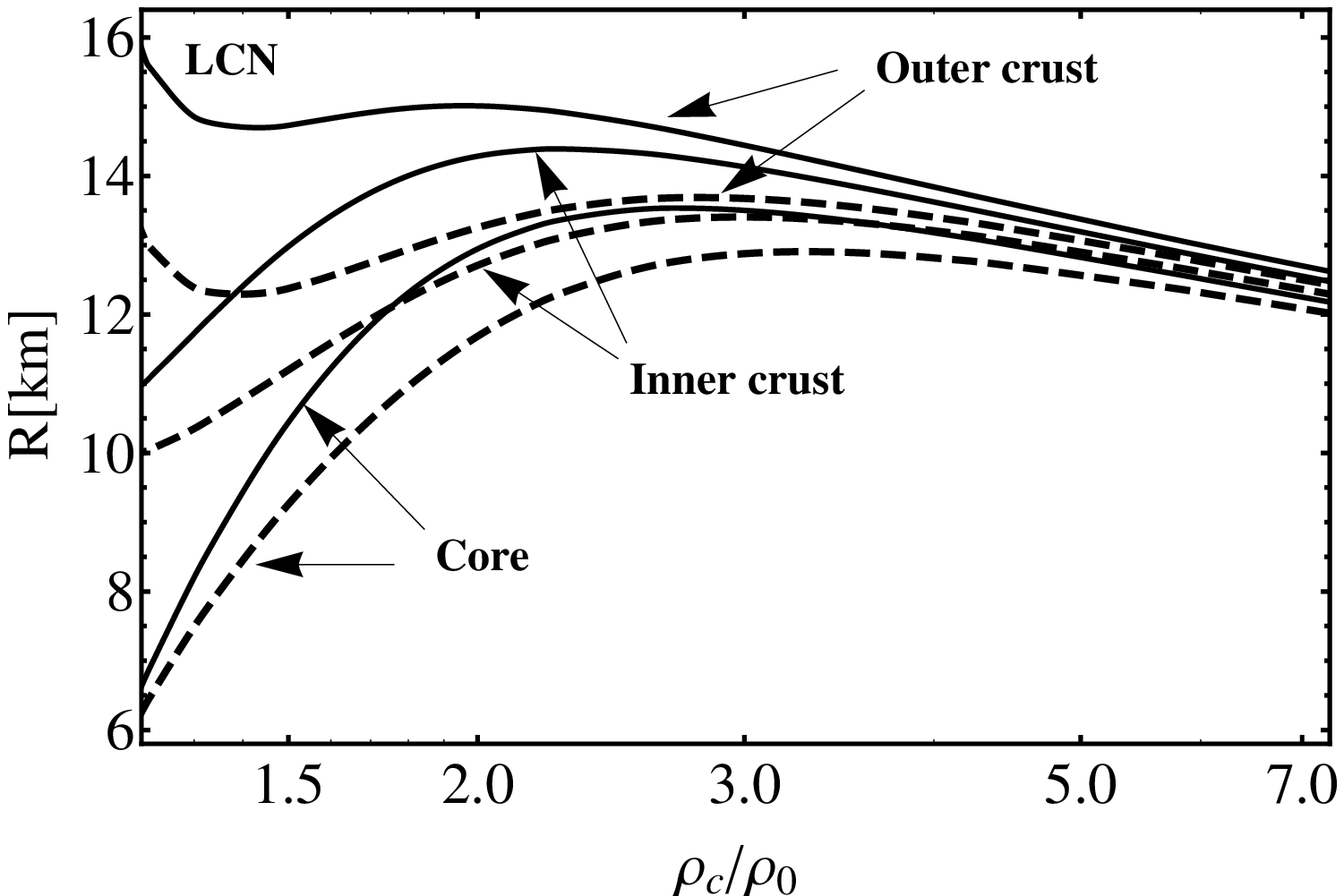}
\caption{Radius versus central density for the locally neutral configuration. All dashed curves are for the static configurations. All solid curves are the equatorial radii for the Keplerian sequence. }\label{fig:lcn_Rrho}
\end{figure*}

In Figs.~\ref{fig:gcn_Rrho} and \ref{fig:lcn_Rrho} we illustrate the dependence of the static and rotating radii for both local and global charge neutrality cases. For small central densities the radius of the core could be smaller than the thickness of the crust. As for larger central densities the core radius increases and the thickness of the crust deacreases.

\subsection{Rotational to gravitational energy ratio}\label{sec:4.5.2}

\begin{figure}[!hbtp]
\centering
\includegraphics[width=0.75\hsize,clip]{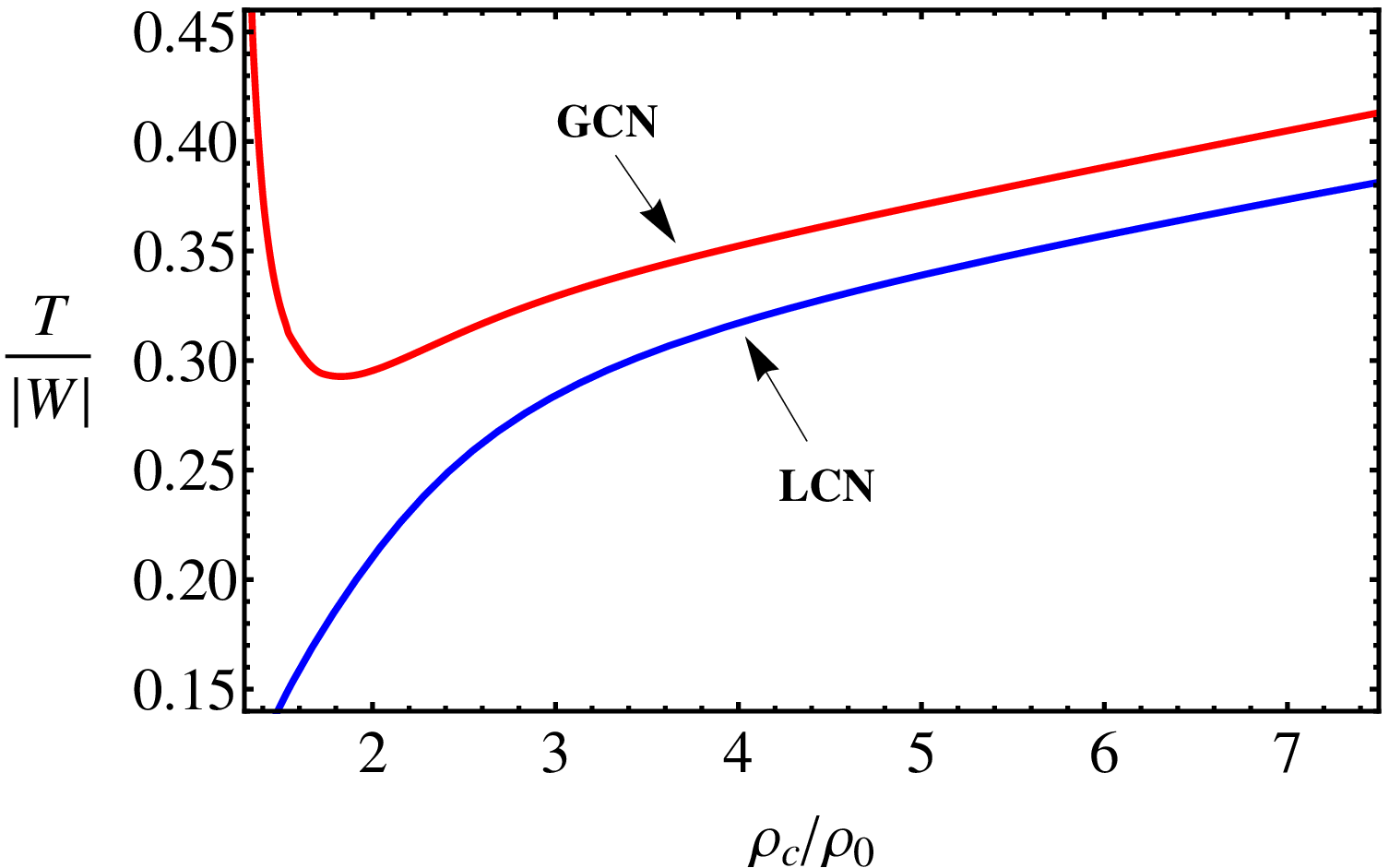}
\caption{Rotational to gravitational binding energy ratio versus central density along the Keplerian sequence both for the global (red) and local (blue) charge neutrality.}\label{fig:ToverWrho}
\end{figure}

Other property of the star related to the centrifugal deformation of the star is the ratio between the gravitational energy and the rotational energy of the star. The former is given by Eq.~(\ref{eq:Wrot}), whereas the latter is
\begin{equation}
T = \frac{1}{2} I \Omega^2+O(\Omega^4),
\end{equation}
In Fig.~\ref{fig:ToverWrho} we plot the dependence of the ratio $T/|W|$ on the central density. As central density increases so does the angular velocity, hence the ratio $T/|W|$ also increases. 

%

\subsection{Angular momentum and quadrupole moment}\label{sec:4.4.3}

\begin{figure*}[!hbtp]
\centering
\includegraphics[width=0.75\columnwidth]{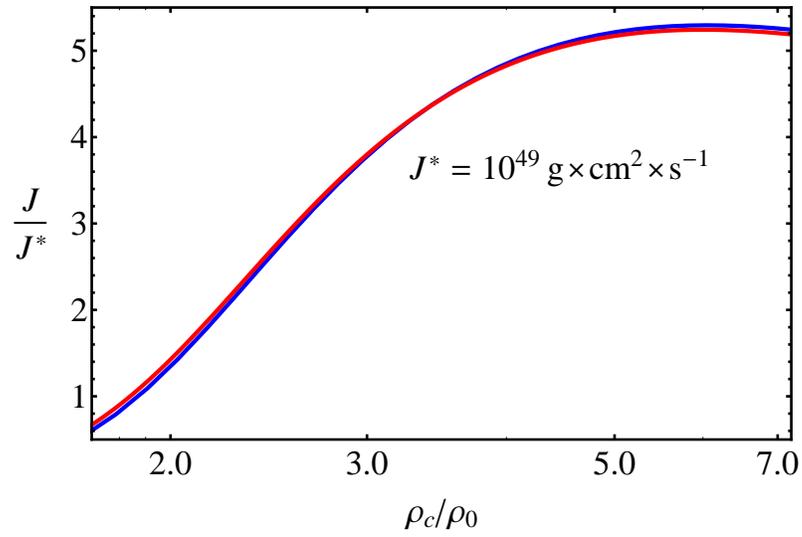}
\caption{Angular momentum as a function of the central density  for both global (red) and local (blue) charge neutrality cases.}\label{fig:glcnJ}
\end{figure*}

\begin{figure*}[!hbtp]
\centering
 \includegraphics[width=0.75\columnwidth]{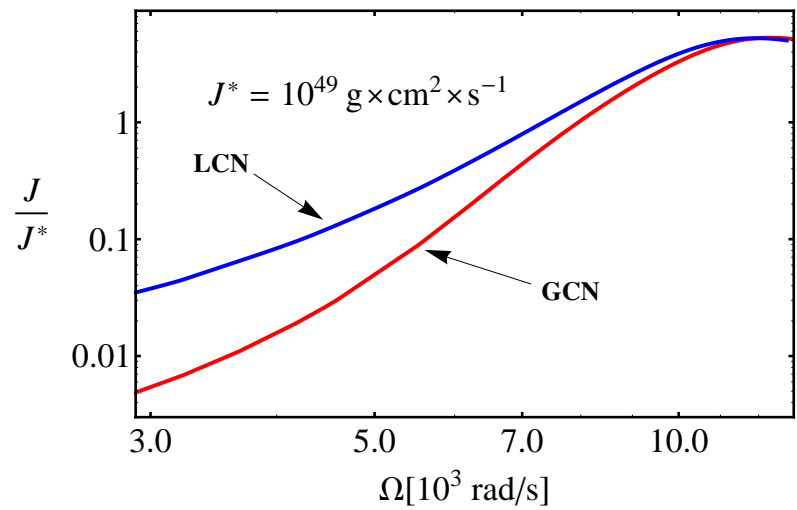}
\caption{Angular momentum versus angular velocity for both global (red) and local (blue) charge neutrality cases. }\label{fig:glcn_JOmg}
\end{figure*}

The angular momentum versus central density is shown in Fig.~\ref{fig:glcnJ}. Here the angular momentum in both global and local neutrality cases are similar. Instead the nonlinear dependence of the angular momentum on the angular velocity in Fig.~\ref{fig:glcn_JOmg}.

\begin{figure}[!hbtp]
\centering
\includegraphics[width=0.75\hsize,clip]{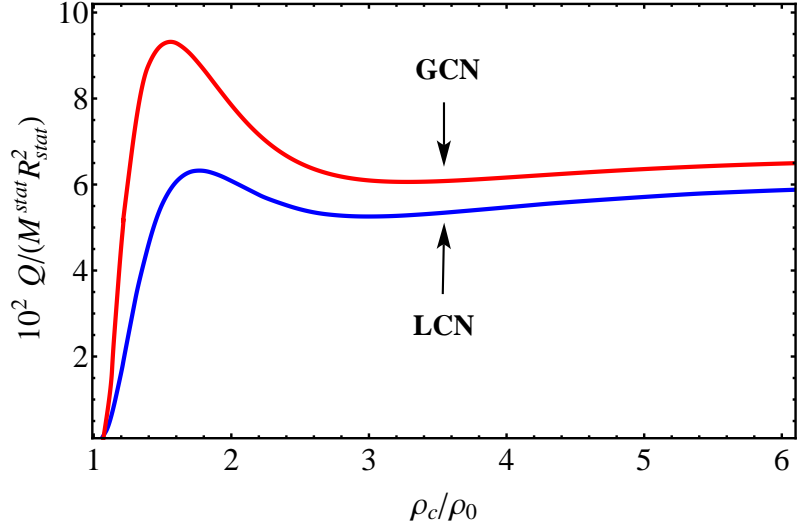}
\caption{Total quadrupole moment versus central density along the Keplerian sequence both for the global (red) and local (blue) charge neutrality cases. The quadrupole moment $Q$ is here in units of the quantity $M R^2$ of the static configuration with the same central density.}\label{fig:Quadrupolerho}
\end{figure}

In Fig.~\ref{fig:Quadrupolerho} we show the quadrupole moment, $Q$ given by Eq.~(\ref{eq:Q}), as a function of the 
central density for both globally and locally neutral neutron stars along the Keplerian sequence.We have normalized the quadrupole moment $Q$ to the quantity $M R^2$ of the non-rotating configuration with the same central density.

\section{Observational constraints}\label{sec:8}

According to observations, the most recent and stringent constraints to the mass-radius relation of neutron stars are provided from data of pulsars by the values of the largest mass, the largest radius, the highest rotational frequency, and the maximum surface gravity. The above mass-radius relations together with the constraints indicated by Tr{\"u}mper (2011) \cite{trumper2011} is shown in Fig.~\ref{fig:constraints}:

\begin{figure}[!hbtp]
\centering
\includegraphics[width=0.75\hsize]{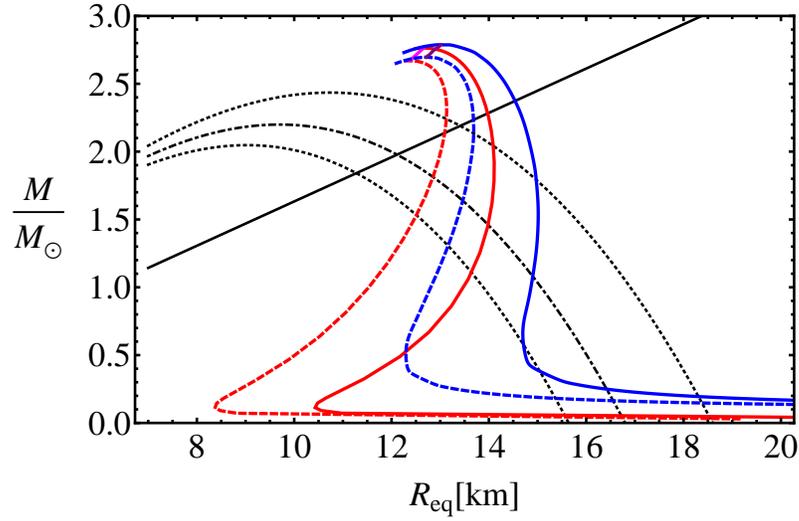}
\caption{Observational constraints on the mass-radius relation given by Ref.~\cite{trumper2011} and the theoretical mass-radius relation presented in Refs.~\cite{2012NuPhA.883....1B,Boshkayev2012thesis,belvedere2014}. The red curves represent the configuration with global charge neutrality, while the blue curves represent the configuration with local charge neutrality. The pink-red line and the light-blue line represent the secular axisymmetric stability boundaries for the globally neutral and the locally neutral case, respectively. The red and blue solid curves represent the Keplerian sequences and the red and blue dashed curves represent the static cases.}\label{fig:constraints}
\end{figure}
%
Up to now the largest neutron star mass measured with a high precision is the mass of the 39.12 millisecond pulsar PSR J0348+0432, $M=2.01 \pm 0.04 M_\odot$ \cite{2013Sci...340..448A}. 

The largest radius is given by the lower limit to the radius of RX J1856-3754, as seen by an observer at infinity $R_\infty = R [1-2GM/(c^2 R)]^{-1/2} > 16.8$ km \cite{trumper04}; it gives the constraint $2G M/c^2 >R-R^3/(R^{\rm min}_\infty)^2$, where $R^{\rm min}_\infty=16.8$ km. 

The maximum surface gravity is obtained by assuming a neutron star of $M=1.4M_\odot$ to fit the Chandra data of the low-mass X-ray binary X7, it turns out that the radius of the star satisfies $R=14.5^{+1.8}_{-1.6}$ km, at 90$\%$ confidence level, corresponding to $R_\infty = [15.64,18.86]$ km, respectively \cite{heinke06}. 

The maximum rotation rate of a neutron star has been found to be $\nu_{\rm max} = 1045 (M/M_\odot)^{1/2}(10\,{\rm km}/R)^{3/2}$ Hz \cite{lattimer2004}. The fastest observed pulsar is PSR J1748-2246ad with a rotation frequency of 716 Hz \cite{hessels06}, which results in the constraint $M \geq 0.47 (R/10\,{\rm km})^3 M_\odot$.

From a technical or practical standpoint, in order to include the above observational constraints in the mass-radius diagram it is convenient to rewrite them for a given range of the radius (for instance, 6 km $\leq$ R $\leq$ 22 km) as follows: \\ 1. The maximum mass:
\begin{equation}\label{maxmass}
\frac{M}{M_{\odot}}=2.01.
\end{equation}
2. The maximum surface gravity:
\begin{equation}
\frac{M}{M_{\odot}}<2.4\times 10^5\frac{c^2}{G}\frac{R}{M_{\odot}}.
\end{equation}
3. The lower limit for the radius surface gravity:
\begin{equation}\label{eq:lowrad}
\frac{M}{M_{\odot}}=\frac{10^5}{2}\frac{c^2}{G}\frac{R}{M_{\odot}}\left(1-\frac{R^2}{(R^{\rm min}_\infty)^2}\right).
\end{equation}
4. In order to include the maximum rotation rate in the rotating mass-radius relation one should construct a constant frequency sequence for the fastest spinning pulsar with 716 Hz. For the sake of generality, we can just require that equilibrium models are bound by the Keplerian sequence (see Refs.~\cite{belvedere2014,cipolletta2015} for details). In all expressions above (\ref{maxmass}-\ref{eq:lowrad}) the mass is normalized with respect to the solar mass $M_{\odot}$ and the radius is expressed in km.

In Fig.~\ref{fig:constraints} we superposed the observational constraints introduced by Tr{\"u}mper~\cite{trumper2011} with the theoretical mass-radius relations presented here and in Belvedere et al. \cite{2012NuPhA.883....1B, belvedere2014} for static and uniformly rotating neutron stars. Any realistic mass-radius relation should pass through the area delimited by the solid black, the dotted-dashed black, the dotted curves and the Keplerian sequences. From here one can clearly see that the above observational constraints show a preference on stiff EoS that provide largest maximum masses for neutron stars. From the above constraints one can infer that the radius of a canonical neutron star of mass $M = 1.4M_{\odot}$ is strongly constrained to $R\geq12$ km, disfavoring at the same time strange quark matter stars. It is evident from Fig.~\ref{fig:constraints} that mass-radius relations for both the static and the rotating case presented here, are consistent with all the observational constraints.

In Table \ref{tab:MRprediction} we show the radii predicted by our mass-radius relation both for the static and the rotating case for a canonical neutron star as well as for the most massive neutron stars discovered, namely, the millisecond pulsar PSR J1614--2230 \cite{demorest2010}, $M=1.97 \pm 0.04 M_\odot$, and the most recent PSR J0348+0432, $M=2.01 \pm 0.04 M_\odot$ \cite{2013Sci...340..448A}.

\begin{table}
\caption{Radii for a canonical neutron star of $M=1.4 M_\odot$ and for PSR J1614--2230 \cite{demorest2010}, $M=1.97 \pm 0.04 M_\odot$, and PSR J0348+0432 \cite{2013Sci...340..448A}, $M=2.01 \pm 0.04 M_\odot$. These configurations are computed under the constraint of global charge neutrality and for a density at the edge of the crust equal to the neutron drip density. The nuclear parametrizations NL3 has been used.}
\centering
\begin{tabular}{c c c}
$M (M_\odot)$ & $R^{J=0}$ (km)	& $R^{J\neq0}_{\rm eq}$ (km)\\
\hline
1.40 & 12.313 & 13.943 \\

1.97 & 12.991 & 14.104 \\

2.01 & 13.020 & 14.097 \\

\hline
\end{tabular}
\label{tab:MRprediction}
\end{table}

Along with the observational constraints one should also take into account the theoretical constraints on the mass-radius relations of neutron stars ~\cite{2016arXiv160607804B}. Indeed only theoretical estimations give the upper and lower bounds for all quantities, describing the properties of neutron stars, to be measured from observations.

\section{Concluding remarks}\label{sec:9}

The equilibrium configurations of static and uniformly rotating neutron stars in both global and local charge neutrality cases have been constructed. To achieve this goal the Hartle approach has been applied to the seed static solution, derived from the integration of the Einstein-Maxwell-Thomas-Fermi equations \cite{2012NuPhA.883....1B}. All physical quantities such as the static and rotating masses, polar and equatorial radii, eccentricity, angular momentum, moment of inertia,  quadrupole moment, rotational kinetic energy and gravitational binding energy have been calculated as functions of the central density and the angular velocity of the neutron star.

In order to investigate only stable configurations of rotating neutron stars the Keplerian mass-shedding limit and the secular axisymmetric instability have been analyzed. This allowed one to construct the stability region, the boundary inside which all stable uniformly rotating neutron stars can be found. In addition the fitting formulas have been obtained for the secular instability boundary in Eqs.~(\ref{eq:SecularG}) and (\ref{eq:SecularL}) for global and local charge neutrality, respectively. With this analysis the maximum mass and maximum rotation frequency of the neutron star have been established. 

In order to favor or disfavor some models of neutron stars the current observational constraints on the mass-radius relations related to the maximum observed mass, maximum surface gravity, largest mass, maximum rotation frequency have been analyzed. All these constraints are of paramount importance not only in the physics of neutron stars, but also in nuclear physics to test theoretical hypothesis and assumptions made in the construction of the equations of state. As a result all observations favor stiff equations of state as indicated in Ref.~\cite{yakovlev2016}.

Finally, the results of this chapter have their immediate astrophysical implications in the physics of compact objects, gravitational waves, short and long gamma ray burst, X-ray phenomena occurring in the accretion disks around neutron stars such as quasi periodic oscillations. Combining both the theory of compact objects and observational data from different phenomena one can infer information not only on the properties and parameters of neutron stars, but also constrain the equations of state, thus probe the nuclear physics theories \cite{2012NuPhA.883....1B, belvedere2014, belvedere2014jkps, 2016arXiv160607804B,yakovlev2016}.

\subsection*{Acknowledgements}

This work was supported by program No F.0679 of grant No 0073 and the grant for the university best teachers-2015 of the Ministry of Education and Science of the Republic of Kazakhstan.


\label{lastpage-01}


\begin{thebibliography}{99}

\bibitem{2013Sci...340..448A}{Antoniadis}, J., {Freire}, P.~C.~C., {Wex}, N., {Tauris}, T.~M., {Lynch},
  R.~S., {van Kerkwijk}, M.~H., {Kramer}, M., {Bassa}, C., {Dhillon}, V.~S.,
  {Driebe}, T., {Hessels}, J.~W.~T., {Kaspi}, V.~M., {Kondratiev}, V.~I.,
  {Langer}, N., {Marsh}, T.~R., {McLaughlin}, M.~A., {Pennucci}, T.~T.,
  {Ransom}, S.~M., {Stairs}, I.~H., {van Leeuwen}, J., {Verbiest}, J.~P.~W.,
  {Whelan}, D.~G., Apr. 2013. {A Massive Pulsar in a Compact Relativistic
  Binary}. Science 340, 448.

\bibitem
  {2012NuPhA.883....1B}
{Belvedere}, R., {Pugliese}, D., {Rueda}, J.~A., {Ruffini}, R., {Xue}, S.-S.,
  Jun. 2012. {Neutron star equilibrium configurations within a fully
  relativistic theory with strong, weak, electromagnetic, and gravitational
  interactions}. Nuclear Physics A 883, 1--24.
  
\bibitem{belvedere2013}
{Belvedere}, R.,  {Rueda}, J.~A., {Ruffini}, R.,
  Jan. 2013. {Neutron Star Cores in the General Relativistic Thomas-Fermi Treatment}. International Journal of Modern Physics: Conference Series 23,  185--192.

\bibitem{belvedere2014}
{Belvedere}, R., {Boshkayev}, K., {Rueda}, J.~A., {Ruffini}, R.,
  Jan. 2014. {Uniformly rotating neutron stars in the global and local charge neutrality cases}. Nuclear Physics A 921, 33--59.
  
\bibitem{belvedere2014jkps}
{Belvedere}, R.,  {Rueda}, J.~A., {Ruffini}, R.,
  Sep. 2014. {Static and rotating neutron stars fulfilling all fundamental interactions}. Journal of the Korean Physical Society 65~(6), 897--902.

\bibitem
  {benhar05}
{Benhar}, O., {Ferrari}, V., {Gualtieri}, L., {Marassi}, S., Aug. 2005.
  {Perturbative approach to the structure of rapidly rotating neutron stars}.
  \prd ~ 72~(4), 044028--+.

\bibitem
{2005MNRAS.358..923B}
{Berti}, E., {White}, F., {Maniopoulou}, A., {Bruni}, M., Apr. 2005. {Rotating
  neutron stars: an invariant comparison of approximate and numerical
  space-time models}. \mnras  ~358, 923--938.

\bibitem
{BBRS2013}
{Bini}, D., {Boshkayev}, K., {Ruffini}, R., {Siutsou}, I., 2013. {Equatorial
  circular geodesics in the Hartle-Thorne spacetime}. Il Nuovo Cimento C 36,
  31.

\bibitem
{boguta77}
{Boguta}, J., {Bodmer}, A.~R., Dec. 1977. {Relativistic calculation of nuclear
  matter and the nuclear surface}. Nuclear Physics A 292, 413--428.

\bibitem
  {2012PhRvD..86f4043B}
{Boshkayev}, K., {Quevedo}, H., {Ruffini}, R., Sep. 2012.
  {Gravitational field of compact objects in general relativity}. \prd ~ 86~(6),
  064043.

\bibitem
  {2012IJMPS..12...58B}
{Boshkayev}, K., {Rotondo}, M., {Ruffini}, R., Mar. 2012. {On
  Magnetic Fields in Rotating Nuclear Matter Cores of Stellar Dimensions}.
  International Journal of Modern Physics Conference Series 12, 58--67.
  
 \bibitem{Boshkayev2012thesis} {Boshkayev}, K. Nov. 2012. {Rotating White Dwarfs and Neutron Stars in General Relativity}. Ph.D. Thesis {\footnotesize \url{ http://padis.uniroma1.it/bitstream/10805/1934/1/Thesis\%20of\%20Kuantay\%20Boshkayev.pdf}}

\bibitem{2016arXiv160607804B} {Boshkayev}, K., {Rueda}, J.A., {Muccino}, M., Jun. 2016. {Theoretical and observational constraints on the mass-radius relations of neutron stars}. ArXiv e-prints: 1606.07804

\bibitem{cipolletta2015} {Cipolletta}, F.  {Cherubini}, C. {Filippi}, S. {Rueda}, J.A. {Ruffini}, R. 2015 {Fast Rotating Neutron Stars with Realistic Nuclear Matter Equation of State}. \prd ~ 92 (2), 023007.

\bibitem
  {demorest2010}
{Demorest}, P.~B., {Pennucci}, T., {Ransom}, S.~M., {Roberts}, M.~S.~E.,
  {Hessels}, J.~W.~T., Oct. 2010. {A two-solar-mass neutron star measured using
  Shapiro delay}. \nat ~ 467, 1081--1083.

\bibitem
  {1988ApJ...325..722F}
{Friedman}, J.~L., {Ipser}, J.~R., {Sorkin}, R.~D., Feb. 1988. {Turning-point
  method for axisymmetric stability of rotating relativistic stars}. \apj  ~325,
  722--724.

\bibitem
  {Friedman1986}
{Friedman}, J.~L., {Parker}, L., {Ipser}, J.~R., May 1986. {Rapidly rotating
  neutron star models}. \apj ~ 304, 115--139.

\bibitem
  {haenselbook}
{Haensel}, P., {Potekhin}, A.~Y., {Yakovlev}, D.~G. (Eds.), 2007. {Neutron
  Stars 1 : Equation of State and Structure}. Vol. 326 of Astrophysics and
  Space Science Library.

\bibitem
{1967ApJ...150.1005H}
{Hartle}, J.~B., Dec. 1967. {Slowly Rotating Relativistic Stars. I. Equations
  of Structure}. \apj ~ 150, 1005.

\bibitem
{1973Ap&SS..24..385H}
{Hartle}, J.~B., Oct. 1973. {Slowly Rotating Relativistic Stars. IX: Moments of
  Inertia of Rotationally Distorted Stars}. \apss ~ 24, 385--405.

\bibitem
{HS1967}
{Hartle}, J.~B., {Sharp}, D.~H., Jan. 1967. {Variational Principle for the
  Equilibrium of a Relativistic, Rotating Star}. \apj ~ 147, 317--+.

\bibitem
{1968ApJ...153..807H}
{Hartle}, J.~B., {Thorne}, K.~S., Sep. 1968. {Slowly Rotating Relativistic
  Stars. II. Models for Neutron Stars and Supermassive Stars}. \apj ~ 153, 807.

\bibitem
  {heinke06}
{Heinke}, C.~O., {Rybicki}, G.~B., {Narayan}, R., {Grindlay}, J.~E., Jun. 2006.
  {A Hydrogen Atmosphere Spectral Model Applied to the Neutron Star X7 in the
  Globular Cluster 47 Tucanae}. \apj ~ 644, 1090--1103.

\bibitem
  {hessels06}
{Hessels}, J.~W.~T., {Ransom}, S.~M., {Stairs}, I.~H., {Freire}, P.~C.~C.,
  {Kaspi}, V.~M., {Camilo}, F., Mar. 2006. {A Radio Pulsar Spinning at 716 Hz}.
  Science 311, 1901--1904.

\bibitem
{klein49}
{Klein}, O., Jul. 1949. {On the Thermodynamical Equilibrium of Fluids in
  Gravitational Fields}. Reviews of Modern Physics 21, 531--533.

\bibitem
  {lalazissis97}
{Lalazissis}, G.~A., {K{\"o}nig}, J., {Ring}, P., Jan. 1997. {New
  parametrization for the Lagrangian density of relativistic mean field
  theory}. \prc ~ 55, 540--543.
  
\bibitem{lattimer2004} {Lattimer} J.M., {Prakash} M. 2004. {The Physics of Neutron Stars} Science, 304, 536.

\bibitem
{oppenheimer39}
{Oppenheimer}, J.~R., {Volkoff}, G.~M., Feb. 1939. {On Massive Neutron Cores}.
  \pr ~ 55, 374--381.

\bibitem{2011PhLB..701..667R}
{Rotondo}, M., {Rueda}, J.~A., {Ruffini}, R., {Xue}, S.-S., Jul. 2011. {The
  self-consistent general relativistic solution for a system of degenerate
  neutrons, protons and electrons in {$\beta$}-equilibrium}. Physics Letters B
  701, 667--671.

\bibitem{2011NuPhA.872..286R}
{Rueda}, J.~A., {Ruffini}, R., {Xue}, S.-S., Dec. 2011. {The Klein first
  integrals in an equilibrium system with electromagnetic, weak, strong and
  gravitational interactions}. Nuclear Physics A 872, 286--295.

\bibitem{shapirobook}
{Shapiro}, S.~L., {Teukolsky}, S.~A., 1983. {Black holes, white dwarfs, and
  neutron stars: The physics of compact objects}.

\bibitem{1981ApJ...249..254S}{Sorkin}, R., Oct. 1981. {A Criterion for the Onset of Instability at a Turning
  Point}. \apj~ 249, 254.

\bibitem{1982ApJ...257..847S}{Sorkin}, R.~D., Jun. 1982. {A Stability Criterion for Many Parameter
  Equilibrium Families}. \apj ~ 257, 847.

\bibitem{2003LRR.....6....3S}{Stergioulas}, N., Jun. 2003. {Rotating Stars in Relativity}. Living Reviews in
  Relativity 6, 3.

\bibitem
{sterfried1995}  {Stergioulas} N., {Friedman} J.L., May 1995 {Comparing models of rapidly rotating relativistic stars constructed by two numerical methods}. \apj ~  444, 306 --311.

\bibitem
{2011MNRAS.416L...1T}
{Takami}, K., {Rezzolla}, L., {Yoshida}, S., Sep. 2011. {A quasi-radial
  stability criterion for rotating relativistic stars}. \mnras ~ 416, L1--L5.

\bibitem
{1930PhRv...35..904T}
{Tolman}, R.~C., Apr. 1930. {On the Weight of Heat and Thermal Equilibrium in
  General Relativity}. Physical Review 35, 904--924.

\bibitem
{tolman39}
{Tolman}, R.~C., Feb. 1939. {Static Solutions of Einstein's Field Equations for
  Spheres of Fluid}. Physical Review 55, 364--373.

\bibitem
{2008AcA....58....1T}
{Torok}, G., {Bakala}, P., {Stuchlik}, Z., {Cech}, P., Mar. 2008. {Modeling the
  Twin Peak QPO Distribution in the Atoll Source 4U 1636-53}. Acta Astronomica
  58, 1--14.

\bibitem
{trumper2011}
{Tr{\"u}mper}, J.~E., Jul. 2011. {Observations of neutron stars and the
  equation of state of matter at high densities}. Progress in Particle and
  Nuclear Physics 66, 674--680.

\bibitem
  {trumper04}
{Tr{\"u}mper}, J.~E., {Burwitz}, V., {Haberl}, F., {Zavlin}, V.~E., Jun. 2004.
  {The puzzles of RX J1856.5-3754: neutron star or quark star?} Nuclear Physics
  B Proceedings Supplements 132, 560--565.

\bibitem
{1992ApJ...390..541W}
{Weber}, F., {Glendenning}, N.~K., May 1992. {Application of the improved
  Hartle method for the construction of general relativistic rotating neutron
  star models}. \apj  ~390, 541--549.
  
  
\bibitem{yakovlev2016} {Yakovlev}, D.G., 2016. {General relativity and neutron stars} International Journal of Modern Physics A ~31 (2 \& 3), 1641017.

\end{thebibliography}
\end{document}